\documentclass[prb,twocolumn,aps,superscriptaddress]{revtex4-2}
\usepackage{graphicx}
\usepackage{dcolumn}
\usepackage{bm}
\usepackage{amssymb}
\usepackage{amsmath}
\usepackage{subfigure}
\usepackage{physics}
\usepackage{sublabel}
\usepackage[utf8]{inputenc}

\usepackage{hyperref}
\usepackage{floatrow}
\usepackage{graphicx}
\usepackage{xcolor}
\usepackage{tabularx}
\usepackage{float}
\floatstyle{plaintop}
\restylefloat{table}

\usepackage{bbm}
\hypersetup{colorlinks = true, linkcolor=blue, citecolor=blue, urlcolor=}
\usepackage[T1]{fontenc}
\begin{document}

%\preprint{APS/123-QED}

\title{On the conclusive detection of Majorana zero modes: conductance spectroscopy, disconnected entanglement entropy and the fermion parity noise
% :\\with Forced Linebreak
}% Force line breaks with \\
%\author[*]{Chaitrali Duse}
%\author[**]{Praveen Sriram}
%\author[***]{Bhaskaran Muralidharan}
%\author[**]{Author D}
%\author[**]{Author E}
%\affil[*]{Department of Physics, Indian Institute of Technology Bombay, Powai, Mumbai-400076, India}
%\affil[**]{Department of Applied Physics, Stanford University, 348 Via Pueblo, Stanford, CA 94305, United States of America}
%\affil[***]{Department of Electrical Engineering, Indian Insititute of Technology Bombay, Powai, Mumbai-400076, India}

%\renewcommand\Authands{ and }
\author{Arnav Arora}
\affiliation{Department of Physics, Indian Institute of Technology Roorkee, Roorkee, Uttarakhand-247667, India}

\author{Abhishek Kejriwal}
\affiliation{Department of Physics, Indian Institute of Technology Bombay, Powai, Mumbai-400076, India}
%  \altaffiliation[Also at ]{Physics Department, XYZ University.}%Lines break automatically or can be forced with \\

\author{Bhaskaran Muralidharan}
\affiliation{Department of Electrical Engineering, Indian Institute of Technology Bombay, Powai, Mumbai-400076, India}
\affiliation{Centre of Excellence in Quantum Information, Computation, Science and Technology, Indian Institute of Technology Bombay, Powai, Mumbai-400076, India}
\email{bm@ee.iitb.ac.in}
 %Authors' institution and/or address

\date{\today}
\begin{abstract}
 Semiconducting nanowires with strong Rashba spin-orbit coupling in the proximity with a superconductor and under a strong Zeeman field can potentially manifest Majorana zero modes at their edges and are a topical candidate for topological superconductivity. However, protocols for their detection based on the local and the non-local conductance spectroscopy have been subject to intense scrutiny. In this work, by taking current experimental setups into account, we detail mathematical ideas related to the entanglement entropy and the fermion parity fluctuations to faithfully distinguish between true Majorana zero modes and trivial quasi-Majorana zero modes. We demonstrate that the disconnected entanglement entropy, derived from the von Neumann entanglement entropy, provides a distinct and robust signature of the topological phase transition which is immune to system parameters, size and disorders. In order to understand the entanglement entropy of the Rashba nanowire system, we establish its connection to a model of interacting spinfull Kitaev chains. Moreover, we relate the entanglement entropy to the fermionic parity fluctuation, and show that it behaves concordantly with entanglement entropy, hence making it a suitable metric for the detection of Majorana zero modes. In connection with the topological gap protocol that is based on the conductance spectra, the aforesaid metrics can reliably point toward the topological transitions even in realistic setups. 

\end{abstract}

\maketitle

%\tableofcontents
%\section{\label{sec:intro}Introduction}
\section{\label{sec:introduction} Introduction} 
Majorana zero modes (MZMs) \cite{kitaev:physusp2001,Aasen-2016,obrien:prl2018,RevModPhys.80.1083} in condensed matter systems have been a crucial talking point, tending to their exotic attributes, such as non-Abelian quantum statistics, which maybe leveraged for futuristic quantum technologies like fault-tolerant topological quantum computation \cite{sarma2015majorana}. Semiconductor-superconductor heterostructures composed of Rashba nanowires proximitised with superconductors are a prime solid-state candidate for realizing MZMs \cite{kitaev:physusp2001,Aasen-2016,obrien:prl2018,RevModPhys.80.1083, doi:10.1063/5.0102999}. Howbeit, conclusive and unambiguous detection of MZMs in these systems has been a matter of intense scrutiny and open deliberation due to controversies involving non-topological origins of MZM-like signatures. Major experimental efforts for the detection MZMs have been based on observing zero-bias conductance peaks (ZBCPs) in the local differential conductance of normal-topological superconductor (N-TS) setups \cite{Das2012,Mourik-2012,Deng-2016,PhysRevLett.110.126406,Scaling_ZBP_Marcus,Albrecht2016,Scaling_ZBP_Marcus}. However, disorder-induced near zero-energy Andreev bound states (ABS), often known as quasi-MZMs, imitate local conductance signatures of MZMs to a great extent \cite{10.21468/SciPostPhys.7.5.061,PhysRevB.86.100503,PhysRevB.97.155425,PhysRevB.96.075161,Lobos,PhysRevB.91.024514,San-Jose2016, PhysRevB.104.075405, PhysRevB.103.214502}, rendering ZBCP based detection of MZMs ambiguous. \\

\begin{figure}[!tbp]
	\centering
    \includegraphics[width=.9\textwidth]{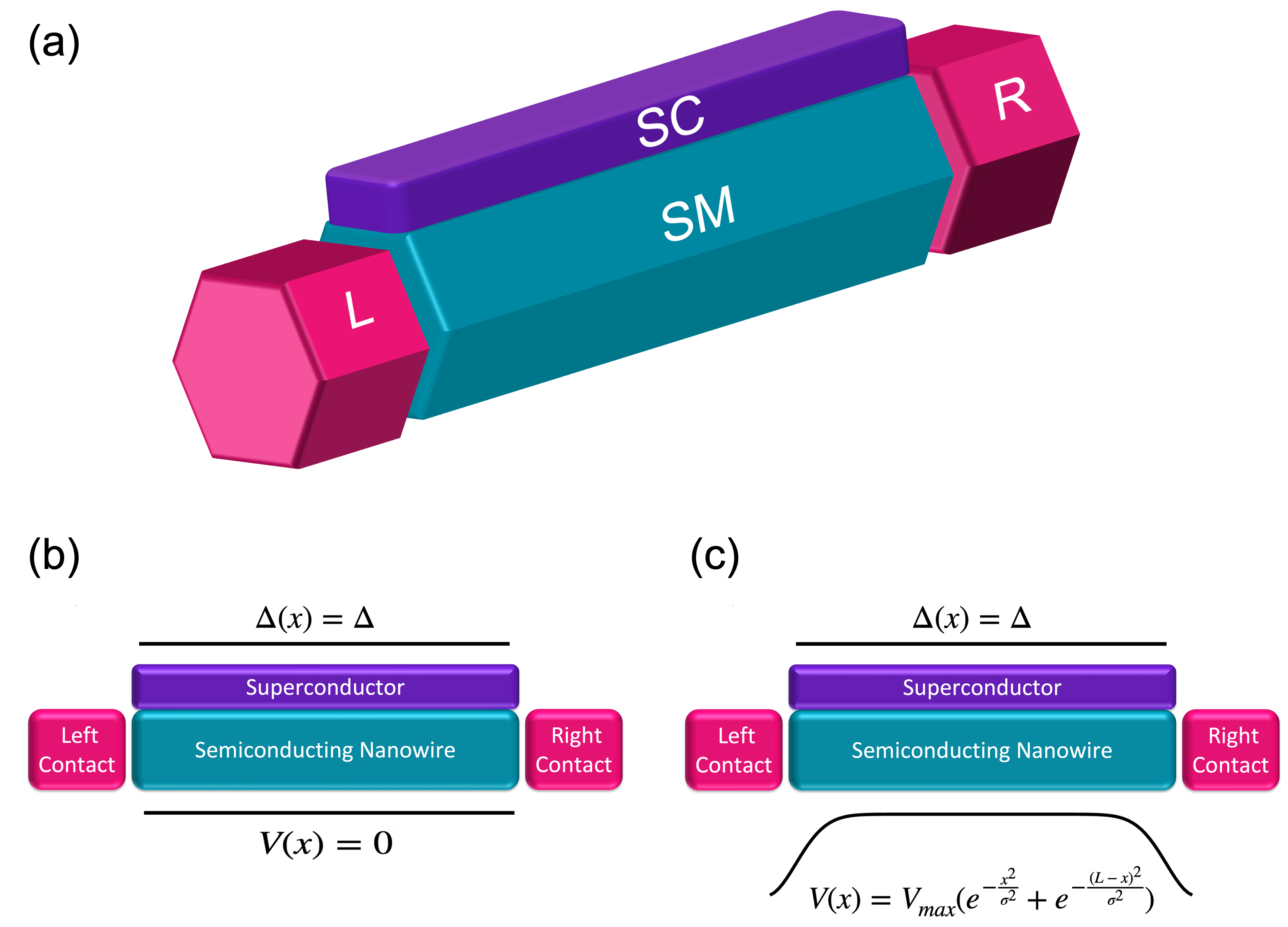}\label{1}
	
	\caption{Device schematics. (a) The N-TS-N system considered in this work. The semiconducting nanowire SM (in blue) is proximitised with a superconductor SC (in purple) which acts as a topological superconductor under a strong spin-orbit coupling and magnetic field. Two normal contacts (in pink) $L$ and $R$ are used to probe the TS system. (b) The pristine setup, with no disorder potential and a homogeneous on-site potential constant throughout the nanowire. (c) The disordered system where the on-site potential varies along the nanowire due to local inhomogeneities.}
 
	\label{fig:1}
\end{figure}
\indent Recent protocols have thus focused on exploiting the non-locality of MZMs through non-local transport spectroscopy using the normal metal-topological superconductor-normal metal (N-TS-N) hybrid setups \cite{PhysRevB.96.195418, PhysRevB.88.180507, PhysRevB.97.045421, puglia, Puglia_Cond_Matrix, Flensberg_Nonlocal, PhysRevB.103.014513, doi:10.1063/5.0102999}. The non-local differential conductance probes the bulk-gap closing and reopening and is presumed to be uninfluenced by the trivial modes \cite{PhysRevB.88.180507}. Ideas proposing measurements of the full conductance matrix, which includes both local and non-local measurements, to identify topological regimes have also been presented \cite{https:MStgp, https://doi.org/10.48550/arxiv.2207.02472}. Nonetheless, numerical \cite{kejriwal2022, Pan_nonlocal} and experimental \cite{puglia} studies have portrayed several issues with the non-local conductance signatures as well, with the signals becoming faint and sputtered \cite{Pan_nonlocal} accompanied by a premature gap closing \cite{kejriwal2022, puglia} in the presence of disorders. This motivates exploring other possibilities that exploit non-local correlations to classify topological orders. One such attribute which can be leveraged to ascertain non-locality in topological phases \cite{PhysRevB.73.245115, PhysRevLett.90.227902, PhysRevLett.109.267203} is the degree of entanglement in the system, which may be quantified using metrics derived from the von Neumann entropy, such as the disconnected entanglement entropy \cite{kejriwal2022}.\\
\indent In this work, we provide an in-depth analysis of various entanglement entropy signatures of MZMs in SM-SC heterostructures. We illustrate that the disconnected entanglement entropy (DEE) \cite{PhysRevB.101.085136} distills out the topological contributions of $ln2$ units to the entanglement entropy \cite{Zeng2019, PhysRevB.101.085136, 10.21468/SciPostPhysCore.3.2.012} and is hence apt for identifying topological phase transitions, unlike the bipartite entanglement entropy (BEE), which is padded with volume and area contributions \cite{kejriwal2022, Bianchi-volume}. In the process, we develop a new pedagogical model, `the spinfull Kitaev chain', to elucidate the entropy signatures of $ln2$ units in Rashba nanowires from a more first-principles perspective. We demonstrate that the entanglement entropy remains robust and quantized over a wide range of controllable parameters and withstands the test of disorder and quasi-MZMs. Further, we inquire into fermion parity noise, a physical observable closely related to entanglement entropy \cite{uncertainity_finland} that maybe be potentially used for its measurement. We conclude by demonstrating the superiority of DEE over the local and non-local conductance spectroscopy \cite{kejriwal2022} for the conclusive detection of MZMs.
\section{\label{sec:methods} Methods} 
The three terminal N-TS-N setup \cite{https:MStgp, https://doi.org/10.48550/arxiv.2207.02472, puglia} that we consider is shown in Fig. \ref{fig:1}(a). The setup comprises a semiconducting nanowire (in blue) with strong Rashba type spin-orbit coupling with an epitaxial layer of s-wave superconductor (in purple), placed in contact with metallic leads `L' and `R' (in pink). The Hamiltonian of the nanowire system in the Bogoliubov-de Gennes (BdG) representation is 
\begin{equation}
\begin{aligned}\label{eq1}
    &\mathcal{H}=\frac{1}{2}\int dx \Psi_{BdG}^{\dagger}(x) H_{BdG} \Psi_{BdG}(x)\\
\end{aligned}
\end{equation}
where $\Psi_{BdG}(x) = (\psi_\uparrow(x), \: \psi_\downarrow(x), \psi_\uparrow^\dagger(x), \: \psi_\downarrow^\dagger(x))^T$ represents the Nambu spinor, and $H_{B d G}$ represents the BdG Hamiltonian \cite{PhysRevB.91.024514} described by:

\begin{equation}
\begin{aligned}\label{eq2}
    &H_{B d G}=\left(-\frac{\hbar^2}{2 m}\partial_x^2 -\mu + V(x)\right) \tau_z \sigma_0+i \alpha_R \tau_z \sigma_y \partial_x\\
    & \; \; \; \; \; \; \; \; \; \;\; \; \; \; \; \; \; \; \; \;\; \; \; \; \; \; \; \; \; \;\; \; \; \; \; \; \; \; \; \; + V_z\tau_z \sigma_x+\Delta \tau_y \sigma_y,
\end{aligned}
\end{equation}
where, $\sigma_i$ are the set of Pauli-matrices in the spin basis, with $\sigma_0$ being the identity matrix and $\tau_i$ are the set of Nambu-matrices in the particle-hole basis. $\mu$ is the electrochemical potential, $\alpha_R$ is the Rashba spin-orbit coupling strength, $V_z = \frac{g\mu_B B}{2}$ is the Zeeman field, $B$ is the magnetic field and $\Delta$ is the proximity-induced superconducting gap. In this work, we elucidate two cases. First, the pristine nanowire case with $V(x) = 0$, portrayed in Fig. \ref{fig:1}(b). Here, both the superconducting gap and the on-site potential are homogeneous and constant throughout the nanowire. As stressed upon earlier, accounting for disorder is crucial for studying conclusive MZM-identification techniques. In the second case, we study the one with disorder where $V(x)$ is non-zero and inhomogeneous. Figure \ref{fig:1}(c) shows the disordered setup where the gap-parameter remains homogeneous but the on-site potential varies along the nanowire due to disorder-induced local inhomogeneities. \\
\indent We analyze smooth inhomogeneous potentials at the ends of the nanowire \cite{Pan-2019-prb, Pan-2020, Pan_nonlocal}. In experiments, such disorders occur as Schottky barrier formations due to Fermi energy mismatches with the contacts \cite{PhysRevB.97.165302} or due to charge inhomogeneities in the environment or the leads \cite{PhysRevLett.123.107703, doi:10.1126/science.aaf3961}.  For our numerical simulations, we model this smooth disorder as a half-Gaussian at each end of the nanowire with a peak value of $V_{max}$ and variance $\sigma$. For our numerical calculations, we discretize \cite{Duse_2021, thesis_superconducting_floquet} the Hamiltonian (\ref{eq1}) which yields a tight-binding Hamiltonian:
 \begin{equation}
 \begin{aligned}\label{H_TB}
    \mathcal{H} &= \sum_i \mathcal{C}_i^\dagger \bigg [\left(2t -\mu + V_i\right) \tau_z \sigma_0 + V_z\tau_z \sigma_x+\Delta \tau_y \sigma_y \bigg ]\mathcal{C}_i \\
    &\; \; \; \; \; \; \; \; \; \;\; \; \; \; \; \; \; \; \; \;\;+ \sum_{\langle ij \rangle} \mathcal{C}_i^\dagger \bigg [-t\tau_z \sigma_0+i t_{SO} \tau_z \sigma_y \bigg ]\mathcal{C}_j,
\end{aligned}
 \end{equation} 
 where, $\mathcal{C}_i = (c_{i\uparrow}, c_{i\downarrow}, c_{i\uparrow}^\dagger, c_{i\downarrow}^\dagger)$ is the Nambu spinor at site `i', $t_{SO} = \alpha_R/2a$, $t = \hbar^2/2ma^2$ and $a$ is the lattice spacing. A direct diagonalization of the tight-binding Hamiltonian \eqref{H_TB} is used to obtain the eigenspectra. \\

\begin{figure}[!tbp]
	\centering
    \includegraphics[width=0.9\textwidth]{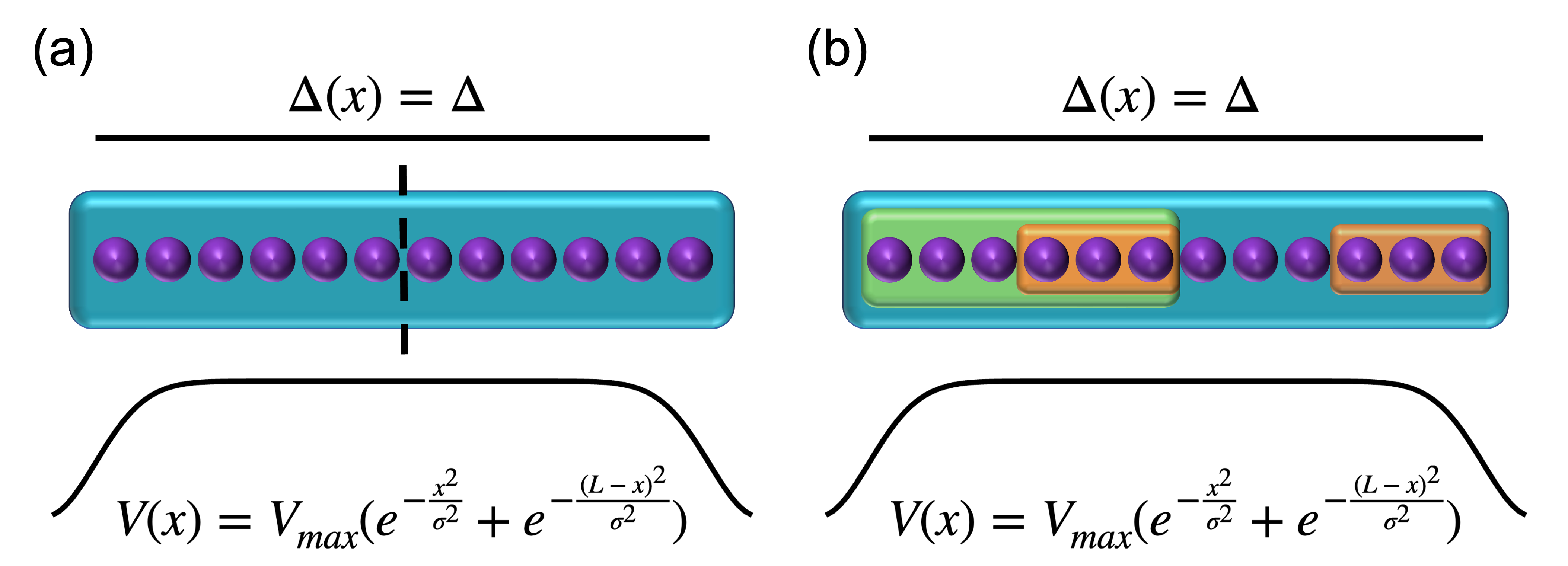}\label{2}
	\caption{ Partitions used for entanglement entropy calculations. (a) A partition in the middle of the wire used for computing the bipartite entanglement entropy. (b) Two distinct partitions, $A$ (in green), and $B$ (in orange), used for computing the disconnected entanglement entropy. The region $A$ is a continuous region formed by partitioning the wire in the middle and $B$ is a disconnected region formed by partitioning the chain into two non-intersecting regions.}
    
	\label{fig:2}
\end{figure}
\textbf{Entanglement Entropy}: For calculating the entanglement entropy, we leverage its intricate connection with the correlation matrix \cite{supple, PhysRevB.73.245115}, which is defined as :
\begin{equation}
    C_{n m}=\left(\begin{array}{cc}\left\langle c^{\dagger} c\right\rangle & \left\langle c^{\dagger} c^{\dagger}\right\rangle \\ \langle c c\rangle & \left\langle c c^{\dagger}\right\rangle\end{array}\right)_{n m}, \end{equation} The matrix elements are calculated using the expectation values of fermionic operators in the BCS ground-state. The truncated correlation matrix of any sub-region of the system, say A, can be defined as:
\begin{equation}
    \Tilde{C}_{n m}=C_{(nm)\in A}
\end{equation}
With a comprehensive calculation \cite{supple}, we can prove that the von Neumann entropy of sub-region A can be expressed in terms of the eigenvalues of the truncated correlation matrix as:
\begin{equation}
    S=-\sum_{k=1}^M\left[\xi_k \log \xi_k+\left(1-\xi_k\right) \log \left(1-\xi_k\right)\right]
\end{equation}
When the system is partitioned into two subsystems, the von Neumann entropy of one of the sub-systems is called the bi-partite entanglement entropy (BEE). In this work, we calculate the BEE by partitioning the system into two equal halves, as shown in Fig. \ref{fig:2}(a). \\
 \indent The disconnected entanglement entropy (DEE) \cite{PhysRevB.101.085136, kejriwal2022} is computed using linear combinations of von Neumann entropy of disconnected regions of the system, an idea similar to the topological entanglement entropy in 2D-systems \cite{TEE-kitaev}. The disconnected regions used in this work can be seen in Fig. \ref{fig:2}(b), we call the green region $A$ and the orange region $B$. The DEE $S_D$ is then calculated as:
\begin{equation}
    S_D = S_A + S_B - S_{A \cap B} - S_{A \cup B} \:\:,
\end{equation}
where, $S_A$, $S_B$, $S_{A \cap B}$, $S_{A \cup B}$ are the von Neumann entropies of the corresponding regions. The DEE dissects out area and volume contributions \cite{kejriwal2022, PhysRevB.101.085136, PhysRevLett.90.227902} to the entanglement entropy hence distilling out only the topological contributions to it. As we will see further, BEE signatures are poised with the issue of volume and area terms but DEE signatures are clean and quantized \cite{kejriwal2022, PhysRevB.101.085136}.  \\
\textbf{Fermion Parity Noise}: We also relate the entanglement entropy to a well known conserved physical observable of the system -- Fermion parity \cite{uncertainity_finland}. The parity operator is given by $\mathcal{P} = (-1)^{\sum_i c_i^\dagger c_i}$. The fermion parity noise is the statistical measurement uncertainty in $\mathcal{P}$ of the BCS ground-state, defined by:

\begin{equation}
    \langle P \rangle_2 = \langle \mathcal{P}^2 \rangle - \langle \mathcal{P} \rangle^2
\end{equation}

where, the expectation value is calculated in the BCS ground-state. It can be shown that the fermion parity noise of any sub-region of a superconducting system can be related to the eigenvalues of the truncated correlation matrix \cite{supple} of that region by:
\begin{equation}
    \langle P \rangle_2 = \sum_{k=1}^M \xi_k(1 - \xi_k)
\end{equation}
We then define disconnected fermion parity noise $\delta P$ on the lines of the DEE, as a linear combination of fermion parity noise of distinct disconnected regions of the nanowire, given by:
\begin{equation}
    \delta P = \langle P_A \rangle_2 + \langle P_B \rangle_2 - \langle P_{A \cap B} \rangle_2 - \langle P_{A \cup B} \rangle_2.
\end{equation}
\textbf{Conductance}: The local and non-local conductance signatures are calculated using the Keldysh non-equilibrium Green's function (NEGF) formalism \cite{kejriwal2022}. This requires evaluating the retarded (advanced) Green's function $G^{r(a)}$ of the system by incorporating the contacts as self energies $\Sigma_{L/R}$ along with the device Hamiltonian $H_{BdG}$. In this work, we use the wide-band approximation to define the broadening matrices $\Gamma_{L/R}$ and subsequently the self energies $\Sigma_{L/R}$ with coupling energy $\gamma = 0.01 \Delta$. The retarded (advanced) Green's functions along with the broadening matrices are then used to calculate the energy resolved Andreev transmission ($T_A(E)$), the crossed-Andreev transmission ($T_{CA}(E)$), and the direct transmission ($T_D(E)$) \cite{supple}, which are consequently used to calculate the local and non-local conductance in the low temperature limit using the Landauer-Büttiker formalism as:
\begin{equation}
\begin{array}{r}
\left.G_{L L}(V)\right|_{T \rightarrow 0} \equiv \frac{e^2}{h}\left[T_A(E=e V)+T_A(E=-e V)+\right. \\
\left.T_{C A R}(E=e V)+T_D(E=e V)\right]
\end{array}
\end{equation}
\begin{equation}
\left.G_{L R}(V)\right|_{T \rightarrow 0} \equiv \frac{e^2}{h}\left[T_D(E=e V) - T_{C A R}(E=-e V)\right],
\end{equation} 
where $V$ is the bias voltage applied at either contact in the three terminal configuration \cite{Puglia_Cond_Matrix}.

\begin{table} 
\caption{Parameters used in the analysis, unless otherwise stated.}
\begin{tabular}{lcc}
\hline Parameter & & Value \\
\hline Effective mass & $m^*$ & $0.015 m_{\mathrm{e}}$ \\
Induced order parameter & $\Delta$ & $0.25 \mathrm{meV}$ \\
Tight-binding hopping parameter & $t_0$ & $10 \mathrm{meV}$ \\
Rashba spin-orbit coupling & $\alpha_{\mathrm{R}}$ & $20 \mathrm{meV} \mathrm{nm}$ \\
Chemical potential & $\mu$ & $0.5 \mathrm{meV}$ \\
Magnetic g-factor & $g$ & 40 \\
Critical field & $B_c$ & $0.58 \mathrm{~T}$ \\
\hline
\end{tabular}
%\label{table1}
\label{table1}
\end{table}

\begin{figure}[!tbp]
	\centering
    \includegraphics[width=1.0\textwidth]{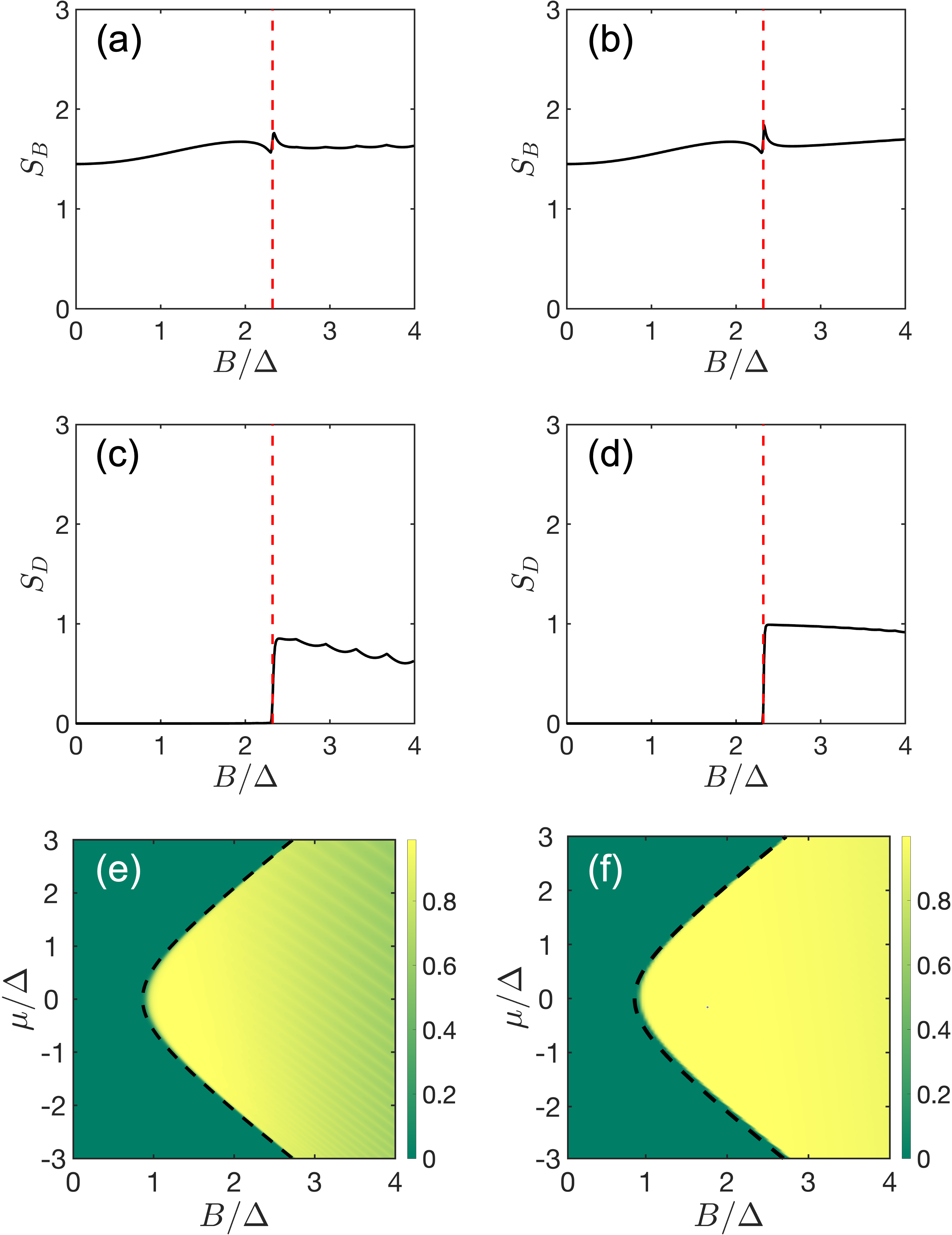}\label{3}

	\caption{Topological phase space and the entanglement entropy. (a) The BEE (in units of $\ln2$) of a $4\mu m$ pristine nanowire. (b) The BEE (in units of $\ln2$) of an $8\mu m$ pristine nanowire. (c) The DEE (in units of $\ln2$) of a $4\mu m$ pristine nanowire. (d) The DEE (in units of $\ln2$) of an $8\mu m$ nanowire. TIn all the cases the entropies are plotted against the magnetic field $B$. The dashed red line is at the theoretically predicted critical field $B_c$ for topological phase transition. The BEE shows no characteristic quantization with a minimal change in its value over the phase transition. The DEE, however, stays close to zero in the trivial regime: $B < B_c$ and is quantized at a value of $\ln2$ in the topological regime: $B > B_c$. (e) DEE phase diagram (in units of ln2) of a $4 \mu m$ nanowire. (f) DEE phase diagram (in units of ln2) of an $8 \mu m$ nanowire.  In (e) and (f) the dashed black curve represents the theoretical phase boundary.  The DEE stays zero in the trivial regime and attains $ln2$ value in the topological regime for a good range of $mu$ and $B$. }
    
	\label{fig:3}
\end{figure}

\begin{figure}[!tbp]
	\centering
    \includegraphics[width=1.0\textwidth]{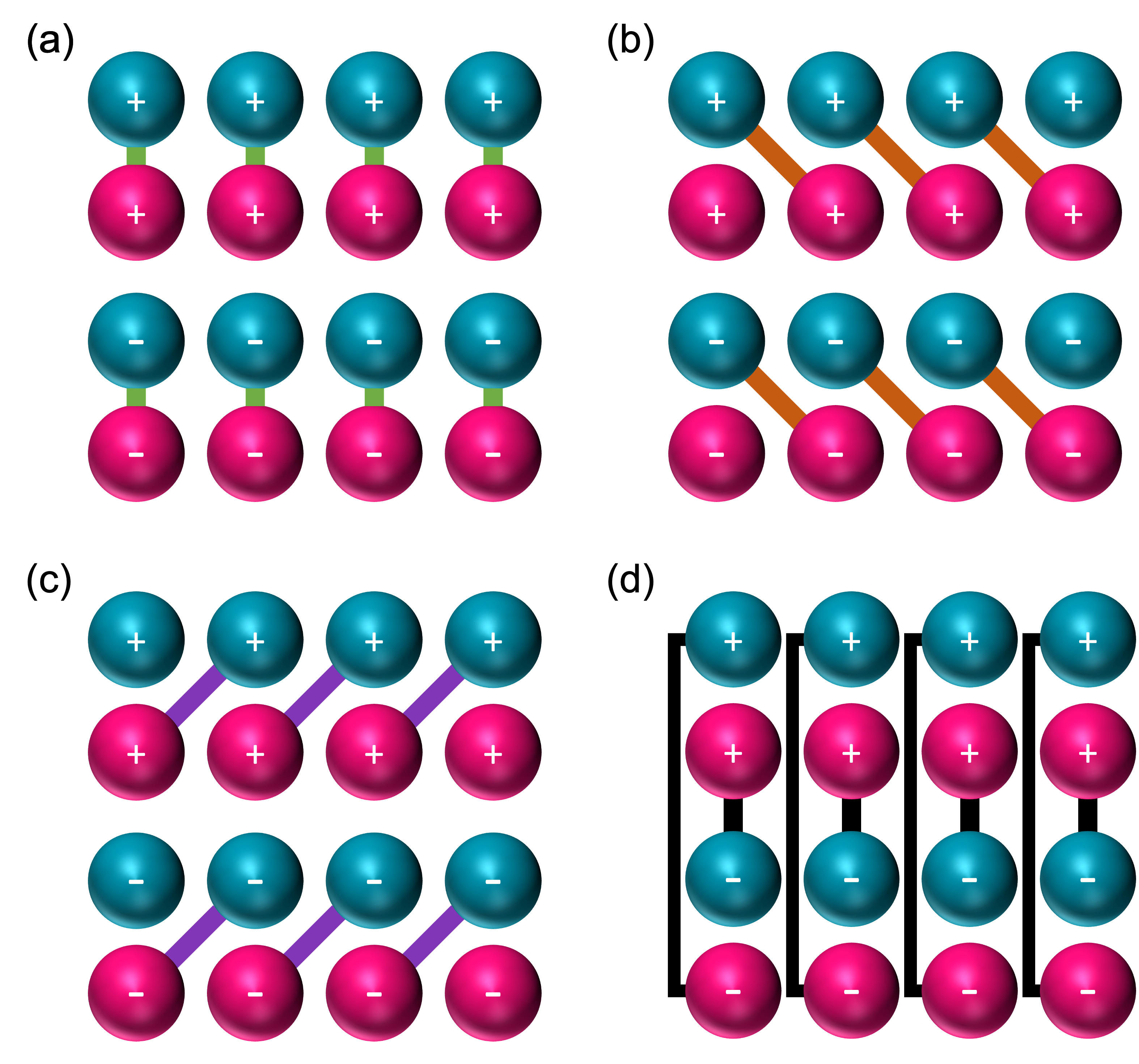}\label{5}

	\caption{ Decomposition of the spinfull Kitaev chain into four interacting Majorana chains. The blue and pink spheres represent $A$ type and $B$ type Majoranas respectively of the $\pm$ chains. (a) the green bonds represent couplings due to $\mu_+$ and $\mu_-$, (b) the orange bonds represent couplings due to $(t_+ - \Delta_+)$ and $(t_- - \Delta_-)$, (c) the violet bonds represent couplings due to $(t_+ + \Delta_+)$ and $(t_- + \Delta_-)$ and (d) the black bonds represent couplings due to $\Delta_s$}
    
	\label{fig:5}
\end{figure}
\section{\label{sec:results} Results}
We present the results obtained from numerical simulations of the system described in Fig. \ref{fig:1} based on the experimentally relevant system parameters presented in Tab. \ref{table1}. We start by calculating the entanglement entropy signatures for the pristine nanowire. Figure \ref{fig:3} focuses on the entanglement entropy as a function of the Zeeman field for short ($4 \mu m$) and long ($8 \mu m$) nanowires. \\
\indent We start by evaluating the BEE signatures for the $4 \mu m$ and $8 \mu m$ nanowires. The BEE shows no signs of quantization and exhibits a minimal change in its value over the phase transition for both $4 \mu m$ and $8 \mu m$ nanowires, as shown in Figs. \ref{fig:3}(a), and \ref{fig:3}(b), respectively, where the red dashed lines mark the critical magnetic field for the topological phase transition. The critical magnetic field $B_c$ is evaluated using the well-known relation: $B_c = 1/g \sqrt{(\mu^2 + \Delta^2)}$. The non-zero values of the BEE in the trivial phase and lack of quantization can be attributed to the area and volume contributions that arise from the bulk of the nanowire over the topological contributions from the ends. These contributions render it ineffective for the identification of the topological phase transition. \\
\indent Figures \ref{fig:3}(c) and \ref{fig:3}(d) present the DEE signatures for the $4 \mu m$ and $8 \mu m$ nanowires respectively. The red dashed line, as earlier, represents $B_c/\Delta$. Unlike the BEE, the DEE remains zero in the trivial regime $B<B_c$. At the critical point, the DEE shows a sudden jump and attains a non-zero quantized value of $ln2$ in the topological regime $B>B_c$. The DEE signature is robust to the system size and shows similar behavior for both short and long nanowires. \\
\indent The quantization towards the value of $ln2$ becomes more prominent for larger system size, as is clear from Fig. \ref{fig:3}(d). We also observe other prominent features of the SM-SC hybrid system in the DEE signatures. The oscillations in the DEE, as seen in Fig. \ref{fig:3}(c), correspond to the well-known parity crossings of the MZMs near zero energy. The amplitude of the parity crossings reduces as the system size increases. This aligns with the reduced oscillatory behavior of the longer nanowire in Fig. \ref{fig:3}(d) than the shorter nanowire in Fig. \ref{fig:3}(c). The near-zero value of the DEE in the trivial regime and a quantized non-zero value in the topological regime, irrespective of system size, indicates its capability to faithfully detect the topological phase transition in the nanowire and distinguish the topological phase from the trivial phase.  \\
\indent In Fig. \ref{fig:3}(e) and Fig. \ref{fig:3}(f), we highlight the robustness of DEE as a metric to distinguish between the topological and trivial phases and detect a topological phase transition by presenting the topological phase diagram for the system. Figure \ref{fig:3}(e) depicts the value of DEE versus an experimentally relevant range of the electrochemical potential and the Zeeman field for a pristine wire of length 4 $\mu m$. The dashed black curve denotes the theoretically evaluated topological phase boundary for a fixed value of $\Delta$ and is defined by the relation $B_c = \frac{1}{g}\sqrt{(\mu^2 + \Delta^2)}$. The region to the right of the curve represents the theoretically expected topological phase. On calculating DEE for each value of $\mu$ and $B$ in this range, we observe that DEE attains a non-zero quantized value of $\ln 2$ in the region to the right of the dashed curve, whereas it stays close to zero to the left of the dashed curve. This shows that the transition in the value of DEE exactly matches the theoretically expected topological phase boundary, and the value of DEE shows a quantization in the topological regime for a wide range of experimentally relevant parameters. A precisely similar feature is observed for a longer nanowire of length $8 \mu m$, as shown in Fig. \ref{fig:3}(f), highlighting the robustness of DEE as a metric to the system size. \\
% \begin{figure}[!tbp]
% 	\centering
%     \includegraphics[width=1.0\textwidth]{4.png}\label{4}

% 	\caption{ The phase diagram of the disconnected entanglement entropy (in units of $\ln 2$) plotted against the magnetic field $B$ and the chemical potential $\mu$ for (a) a $4\mu m$ pristine nanowire and (b) an $8\mu m$ pristine nanowire. The dashed black curve represents the theoretically calculated topological phase boundary. The DEE successfully detects phase transitions for a reasonable range of experimentally controllable parameters $\mu$ and $B$}
    
% 	\label{fig:4}
% \end{figure}

\begin{figure}[!tbp]
	\centering
    \includegraphics[width=1.0\textwidth]{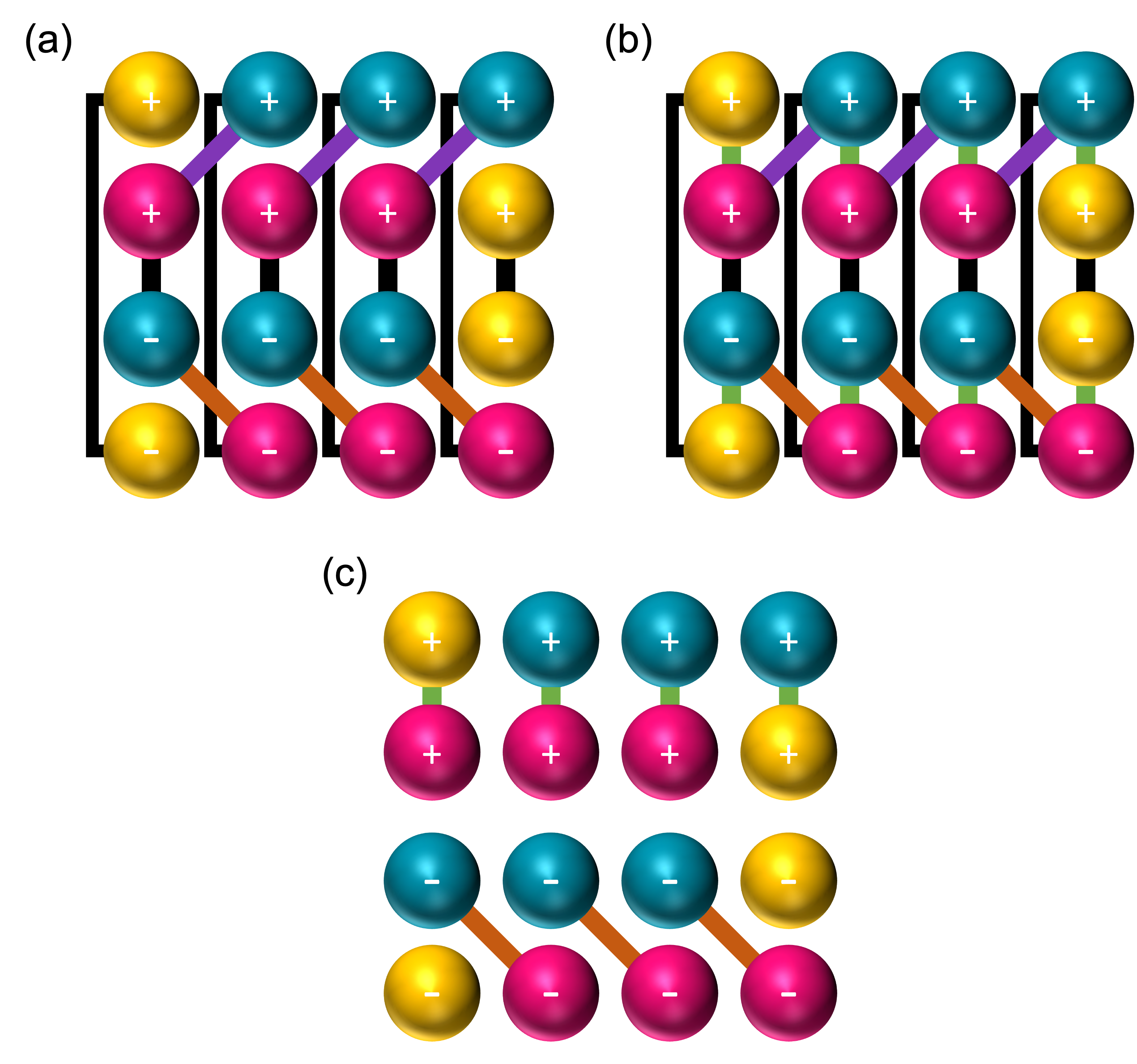}\label{6}
	\caption{ Three different cases considered for the spinfull Kitaev chain. (a) The initial case with $\mu_+ = 0$. In this case the unpaired Majoranas at the end of both the chains (yellow spheres) hybridise through the inter-chain bonding (black bonds) due to the s-wave like pairing between the chains. (b) The intermediate case with $0 < \mu_+ < \infty$. In this case, intra-chain pairing (green bonds) between Majoranas of the $+$ chain (top chain) has been seeded. (c) The final case with $\mu_+ = \infty$. In this case bonds due to $\mu_+$ dominate the $+$ chain making it trivial and isolating it from any other intra-chain or inter-chain bonds. The bottom chain, however, remains topological with unpaired Majoranas at the ends. Overall the system has two MZMs at the ends.   }
    
	\label{fig:6}
\end{figure}
\indent In order to explain the entanglement entropy signatures for the Rashba nanowire from a much more elementary and first-principles perspective, we we highlight the parallels between the Rashba nanowire and a toy model consisting two distinct p-wave superconductors interacting through an s-wave pairing potential. Starting with the normal part of the Rashba nanowire Hamiltonian, without proximity-induced superconductivity, in the continuous momentum space:
\begin{equation}
    \mathcal{H}_0=\int dk \Psi^{\dagger}(k) \bigg[\frac{-\hbar^2 k^2}{2 m}-\mu+i\alpha_R k \sigma_y +V_z \sigma_x\bigg] \Psi(k),
\end{equation}
where, $\alpha_R$ is the Rashba spin-orbit coupling strength, $V_z = \frac{g\mu_BB}{2}$ is the Zeeman field, $k$ is the momentum and $\Psi(k) = (\psi_\uparrow(k), \: \psi_\downarrow(k))^T$ is the fermionic spinor at $k$. The eigenvalues $\varepsilon_\pm$ of this Hamiltonian are given by $\varepsilon_{k, \pm}=\frac{\hbar^2 k^2}{2 m}-\mu \pm \sqrt{V_z^2+\alpha_R^2 k^2}$. The eigen-spinors in terms of the original spinor are given by:
\begin{equation}
\left(\begin{array}{c}
    \psi_+(k)\\
    \psi_-(k)
\end{array}\right)
= 
\left(\begin{array}{cc}
    \frac{\left(-i \alpha_R k+V_z\right)}{\sqrt{V_z^2+\alpha_R^2 k^2}} & 1 \\
    \frac{\left(i \alpha_R k-V_z\right)}{\sqrt{V_z^2+\alpha_R^2 k^2}} & 1 
\end{array}\right)
\left(\begin{array}{c}
    \psi_\uparrow(k)\\
    \psi_\downarrow(k)
\end{array}\right).
\end{equation}
The spinor basis, defined by $\Phi^{'} = (\psi_+(k), \: \psi_-(k))^T$ is often called the helical basis. Now we express the full Hamiltonian of the nanowire, including the proximity-induced superconducting pairing terms, in the helical basis. The Hamiltonian in the helical basis \cite{San-Jose_2013} is given by :
\begin{equation}
\begin{aligned}
    \mathcal{H} &= \frac{1}{2} \int dk [\varepsilon_+(k) \psi^\dagger_+(k)\psi_+(k) + \varepsilon_-(k) \psi^\dagger_-(k)\psi_-(k)] \\ 
    & + [\frac{\Delta_{++}(k)}{2} \psi^\dagger_+(k)\psi^\dagger_+(-k) + \frac{\Delta_{--}(k)}{2} \psi^\dagger_-(k)\psi^\dagger_-(-k) \\ 
    & + \Delta_{+-}\psi^\dagger_+(k)\psi^\dagger_-(-k) + h.c.]
\end{aligned}
\end{equation}
where, 

\begin{equation}
\begin{aligned}
    &\varepsilon_\pm(k) =\frac{\hbar^2 k^2}{2 m}-\mu \pm \sqrt{V_z^2+\alpha_R^2 k^2} \\
    &\Delta_{--,++}(k) = \frac{\pm i \alpha_R k \Delta}{\sqrt{V_z^2+\alpha_R^2 k^2}}\\ 
    &\Delta_{+-}(k) = \frac{V_z \Delta}{\sqrt{V_z^2+\alpha_R^2 k^2}}
\end{aligned}
\end{equation}

\begin{figure}[!tbp]
	\centering
    \includegraphics[width=1.0\textwidth]{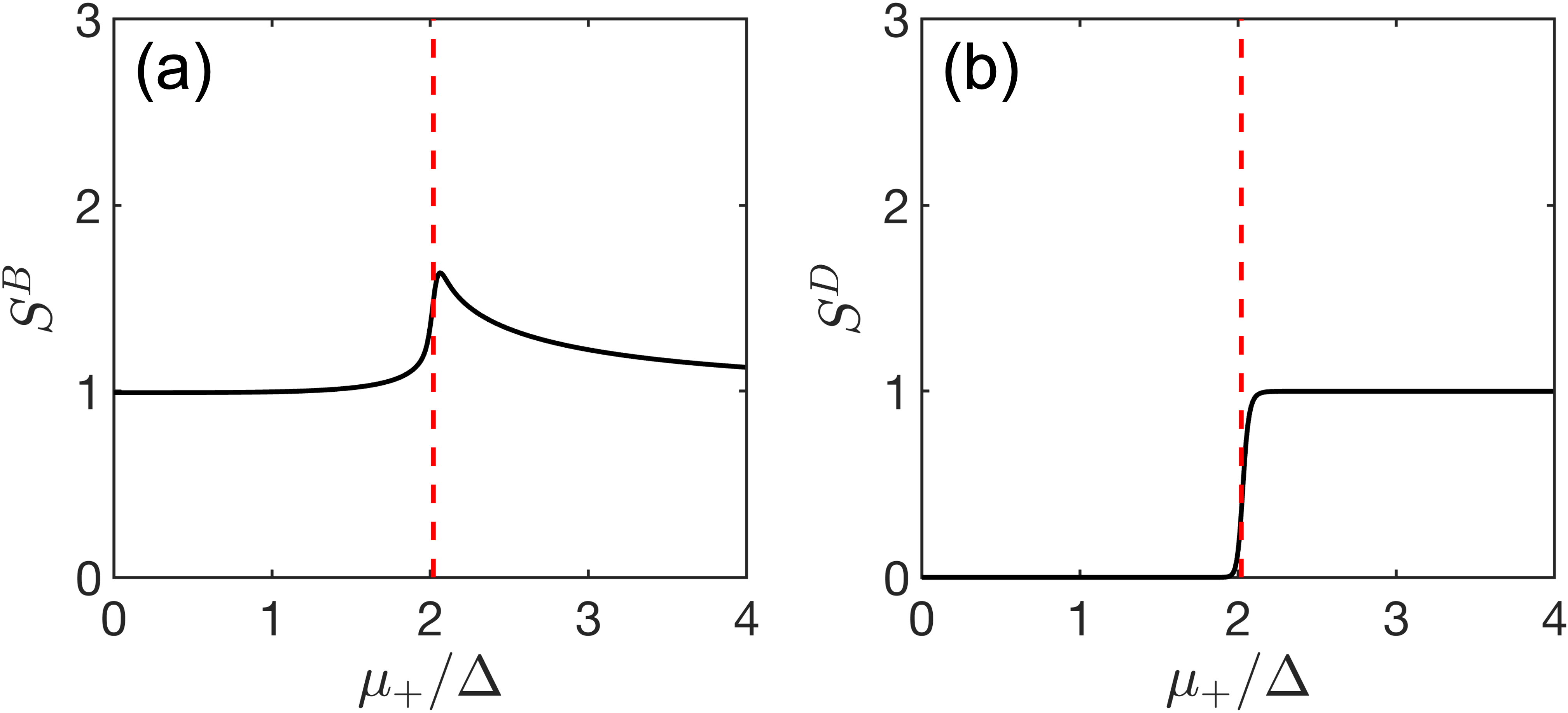}\label{7}
	\caption{ Distillation of the size effects in the spinfull Kitaev model. (a) The BEE (in units of $\ln2$) and (b) the DEE (in units of $\ln2$) of a $100$ site spinfull Kitaev chain plotted against the chemical potential of the $+$ chain $\mu_+$. The spinfull Kitaev chain, a pedagogical model of the nanowire system, develops similar entropy signatures to the nanowire system explaining the characteristics of the BEE and DEE plots of the nanowire from elementary principles. }
    
	\label{fig:7}
\end{figure}

\begin{figure}[!tbp]
	\centering
    \includegraphics[width=1.0\textwidth]{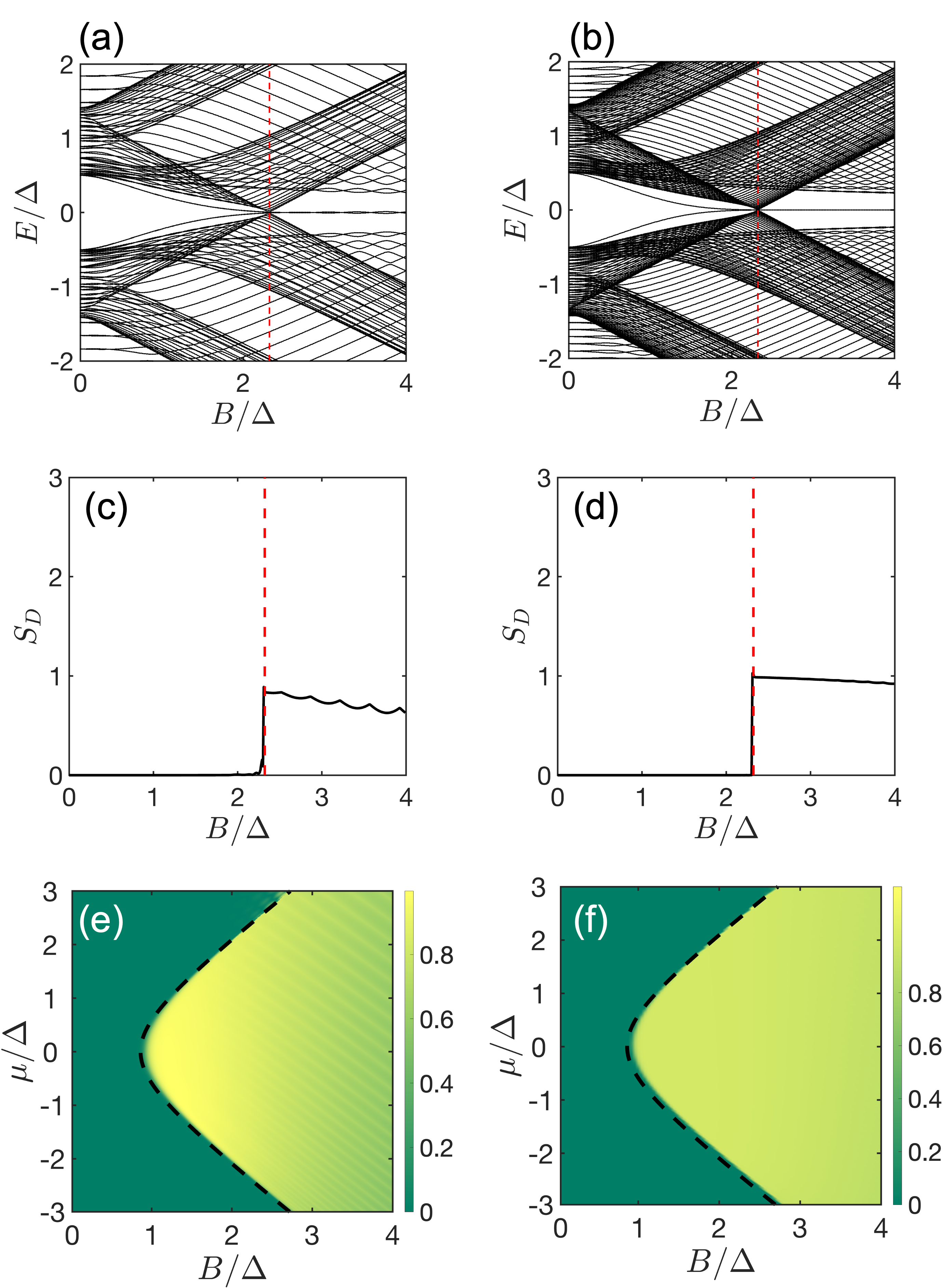}\label{8}
	\caption{Energy spectra, the DEE and the topological phase space. (a) The energy dispersion (in units of $\Delta$) of a $4 \mu m$ disordered nanowire, (b) the energy dispersion (in units of $\Delta$) of an $8 \mu m$ disordered nanowire, (c) DEE of the $4 \mu m$ disordered nanowire and (d) DEE of the $8 \mu m$ disordered nanowire plotted against the magnetic field. (e) The DEE phase diagram (in units of $ln2$) for the $4 \mu m$ disordered nanowire and (f) The DEE phase diagram (in units of $ln2$) for the $8 \mu m$ disordered nanowire. Near zero-energy states before the topological phase transition (red dashed line) are visible in (a) and (b). Neverthless, DEE remains faithfull, rising to $ln2$ units only in the topological regime for both short (c) and long (d) nanowires. DEE also remains robust to varying system parameters, even in the presence of disorder, as seen through the phase diagram (e) and (f)}
    
	\label{fig:8}
\end{figure}
 The Hamiltonian in \eqref{H_TB} can be interpreted as follows: there are two bands $+$ and $-$ with their normal parts being described by $\varepsilon_+(k)$ and $\varepsilon_+(k)$ respectively. Within each band, there exists an intra-band superconducting pairing given by $\Delta_{--, ++}(k)$. These pairings are an odd function of $k$ and hence manifest as intra-band p-wave pairings. There also exists an inter-band superconducting pairing between the two bands given by $\Delta_{+-}(k)$. This pairing is an even function of $k$ and manifests as an inter-band s-wave pairing. Thus, the helical basis allows us to decompose the nanowire into two p-wave superconductors $+$ and $-$ interacting via an s-wave coupling. We notice two things here. First, the effective onsite potential $\mu_{eff} = \mu \mp \sqrt{V_z^2+\alpha_R^2 k^2}$ depends on a controllable parameter $V_z$ and therefore the magnetic field $B$. It decreases with $B$ for the $+$ chain while increases with $B$ for the $-$ chain. Second, the p-wave pairings $\Delta_{--,++}$ are of opposite parity for both bands.\\
\indent This motivates introducing a new pedagogical model, which we call the "spinfull Kitaev chain" based on a topical p-wave superconductor -- The Kitaev chain. We consider two distinct Kitaev chains, labelled with $+$ and $-$, interacting via an s-wave pairing potential $\Delta_s$. The chemical potentials, hopping terms, and p-wave pairings are represented by $\mu_\sigma$, $t_\sigma$, $\Delta_\sigma$, where $\sigma = + \text{ or } -$. The Hamiltonian, describing this pedagogical model, in the position space (with open boundary conditions) is given by:

\begin{figure}[!tbp]
	\centering
    \includegraphics[width=1.0\textwidth]{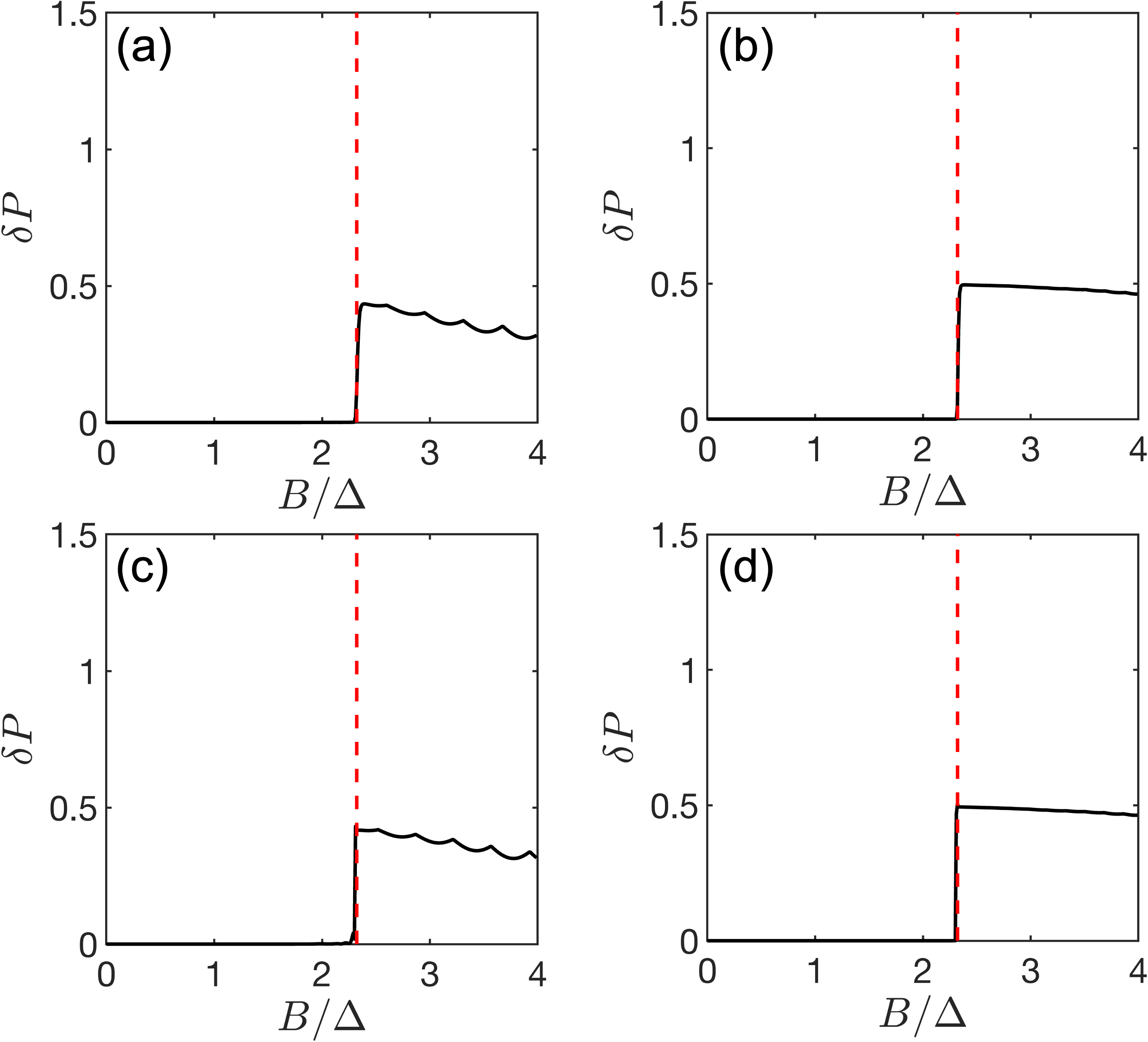}\label{11}
	\caption{ The disconnected fermion parity noise of the nanowire plotted against the magnetic field $B$ for (a) a 4$\mu m$ pristine nanowire, (b) an 8$\mu m$ pristine nanowire, (c) a 4$\mu m$ disordered nanowire and (d) an 8$\mu m$ disordered nanowire. The fermion parity noise of disconnected regions behaves similar to the DEE, with zero value in the trivial regime and $0.5$ quantization in the topological regime both in pristine as well as disordered setups.}
    
	\label{fig:11}
\end{figure}
\begin{equation}
\begin{aligned}
    \mathcal{H} &= \sum_{i=1}^N \sum_{\sigma=\pm}\mu_\sigma c_{i, \sigma}^\dagger\: c_{i, \sigma} \\
    & + \sum_{i=1}^{N-1} \sum_{\sigma=\pm}  t_\sigma c_{i+1, \sigma}^\dagger\: c_{i, \sigma} + h.c.\\   
    & + \sum_{i=1}^{N-1} \sum_{\sigma=\pm} \Delta_\sigma c_{i+1, \sigma}^\dagger\: c_{i, \sigma}^\dagger + h.c.\\
    & + \sum_{i=1}^N \Delta_s c_{i, +}^\dagger\: c_{i, -}^\dagger + h.c. ,
\end{aligned}
\end{equation}
where $c_{i,\sigma}^\dagger$ creates a fermion at site `i' in the $\sigma = \pm$ chain. For further analysis of the model, we express our Hamiltonian in the Majorana basis defined by:
\begin{equation}
    \begin{aligned}
        &c_{i, \sigma} =\gamma_{i, \sigma, A}+i \gamma_{i, \sigma, B} \\
        &c_{i, \sigma}^\dagger =\gamma_{i, \sigma, A}-i \gamma_{i, \sigma, B} \\
        &\gamma_{i, \sigma, \tau}^\dagger = \gamma_{i, \sigma, \tau} \\
        &\{ \gamma_{i, \sigma_i, \tau_i} , \gamma_{j, \sigma_j, \tau_j} \} = 2\: \delta_{i,j} \: \delta_{\sigma_i,\sigma_j}\: \delta_{\tau_i, \tau_j}
    \end{aligned}
\end{equation}
where $i$ represents the site index, $\sigma = + \text{ or } -$ and $\tau = A \text{ or } B$, where $A$ and $B$ represent two distinct types of Majoranas that constitute a single fermion. In the Majorana basis, our model comprises four interacting Majorana chains. The Hamiltonian in this basis is given by:

\begin{figure}[!tbp]
	\centering
    \includegraphics[width=1.0\textwidth]{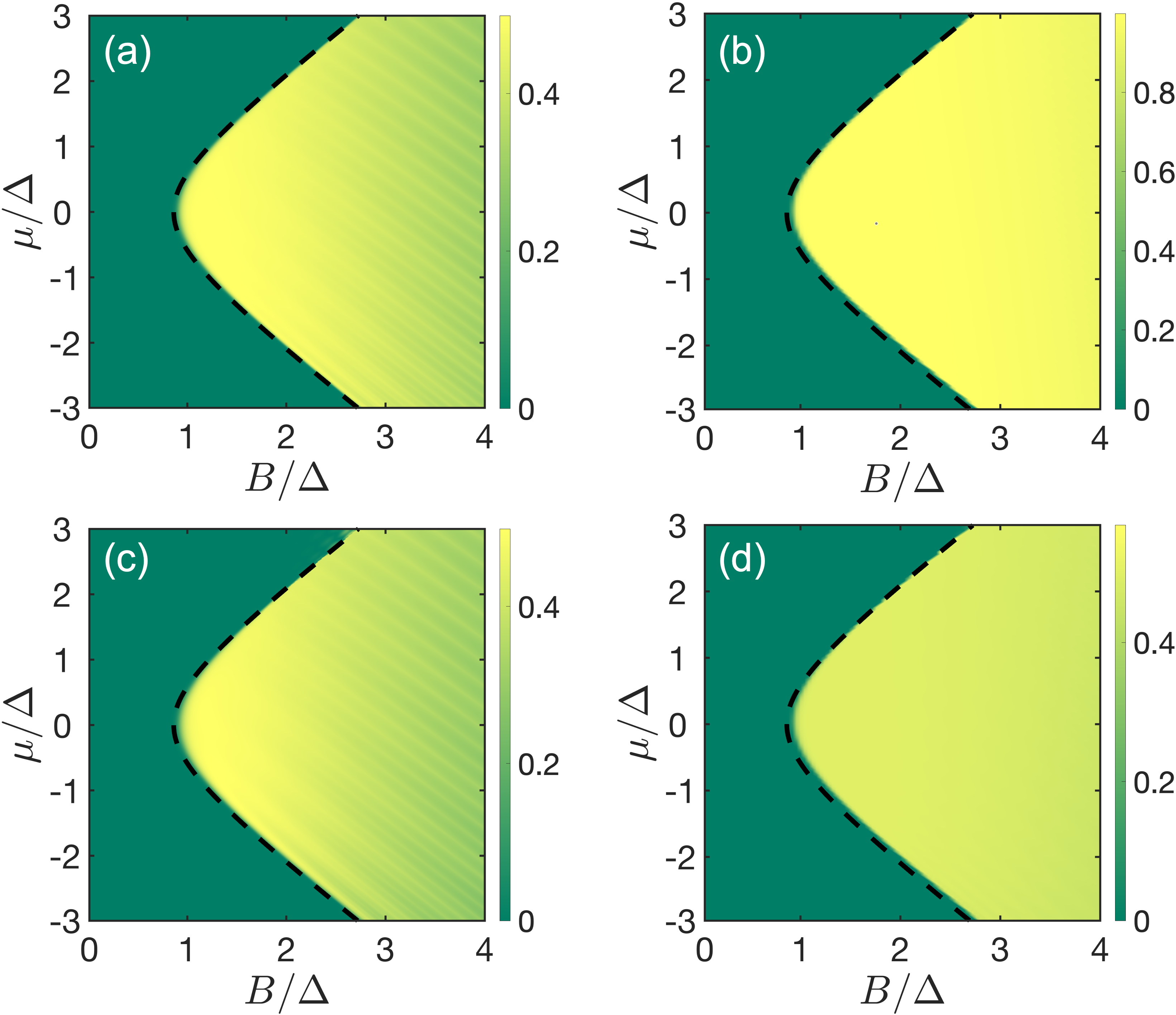}\label{12}

	\caption{Phase diagram of fermion parity noise of disconnected regions of the nanowire plotted against the magnetic field $B$ for (a) a 4$\mu m$ pristine nanowire, (b) an 8$\mu m$ pristine nanowire, (c) a 4$\mu m$ disordered nanowire and (d) an 8$\mu m$ disordered nanowire. The fermion parity noise of disconnected regions faithfully detects topological phase transitions for a good range of $B$ and $\mu$}
    
	\label{fig:12}
\end{figure}
\begin{equation}
\begin{aligned}
H &=\sum_{i=1}^N \sum_{\sigma=\pm}\mu_{\sigma} +\mu_{\sigma} \: \gamma_{i, \sigma, A} \: \gamma_{i, \sigma, B}\\
&+\sum_{i=1}^{N-1}\sum_{\sigma=\pm}\left(t_{\sigma}+\Delta_{\sigma}\right) \gamma_{i+1, \sigma, A} \: \gamma_{i, \sigma, B}\\
&+\sum_{i=1}^{N-1}\sum_{\sigma=\pm}\left(t_{\sigma}-\Delta_{\sigma}\right) \gamma_{i, \sigma, A} \: \gamma_{i+1, \sigma, B} \\
&+\sum_{i=1}^N\Delta_s \: \gamma_{i, -, A} 
\: \gamma_{i, +, B}+\Delta_s \: \gamma_{i, -, B} \: \gamma_{i, +, A}. 
\end{aligned}
\end{equation}
In Fig. \ref{fig:5}, we show the decomposition of the spinfull Kitaev chain into four Majorana chains and show the various pairings of Majoranas caused by different terms of the Hamiltonian as solid bonds. In all the figures, the blue spheres with $+$ represent $\gamma_{i, +, A}$, the pink spheres with $+$ represent $\gamma_{i, +, B}$, the blue spheres with $-$ represent $\gamma_{i, -, A}$, the pink spheres with $-$ represent $\gamma_{i, -, B}$. The green bonds in Fig. \ref{fig:5}(a) represent the terms  $\mu_+$ and $\mu_-$, the orange bonds in Fig. \ref{fig:5}(b) represent the terms  $(t_+ - \Delta_+)$ and $(t_- - \Delta_-)$, the violet bonds in Fig. \ref{fig:5}(c) represent the terms  $(t_+ + \Delta_+)$ and $(t_- + \Delta_-)$, the black bonds in Fig. \ref{fig:5}(d) represent the term  $\Delta_s$. \\

\begin{figure*}[!tbp]
	\centering
    \includegraphics[width=1.0\textwidth]{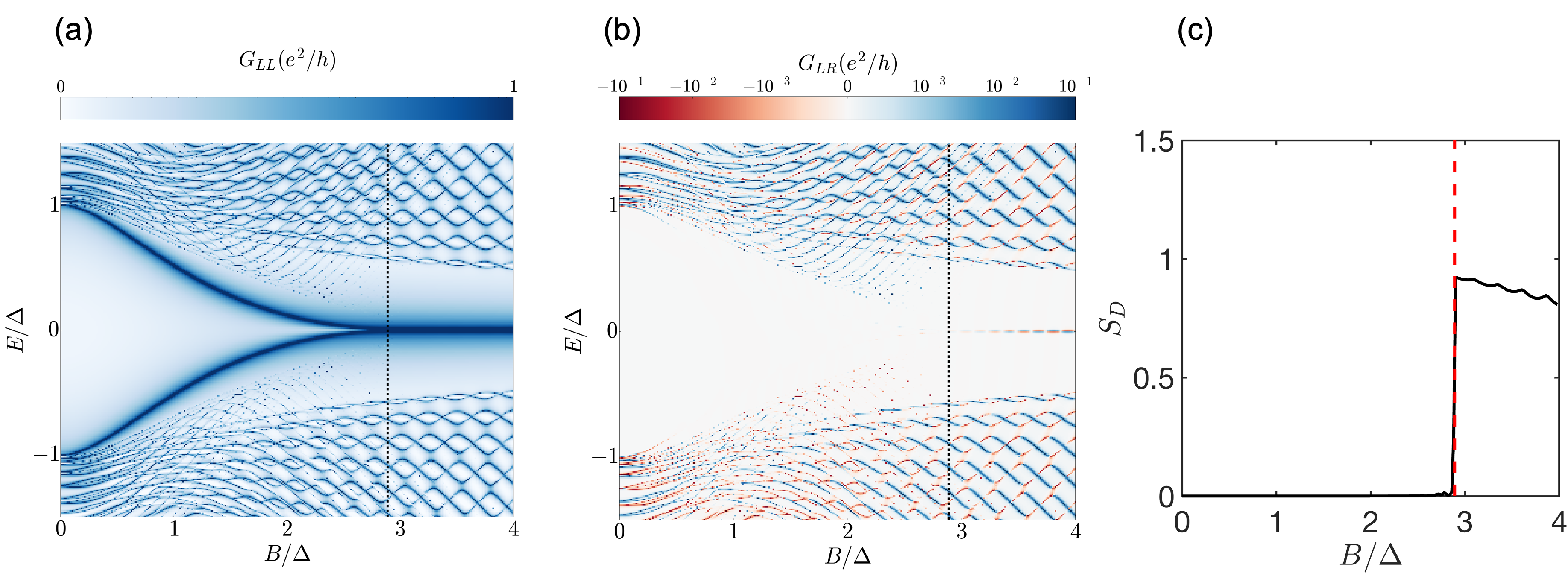}\label{13}

	\caption{ Conductance spectroscopy and the DEE. A comparison of (a) the local conductance, (b) the non-local conductance and (c) DEE of an $6 \mu m$ intermediate length nanowire. The chemical potential is kept at $3.2 \Delta$ and maximum value of disorder is kept at $1 \Delta$. Even under a relatively minimal value of disorder, the local conductance closes before the critical field (black dashed line in (a)). The non-local conductance becomes faint and signature becomes obscure. Nonethless, the DEE remains faithful -- signaling the transition with pinpoint accuracy.}
    
	\label{fig:13}
\end{figure*}

Next, we study the spinfull Kitaev chain by setting $\mu_-=0$, $\Delta_+ = -\Delta_- = 1$, $t_+ = t_- = 1$, $\Delta_s = 0.2$ and we keep $\mu_+$ as the varying parameter. The motivation of varying $\mu_+$ comes from the nanowire model where the $+$ band's effective potential is varied through $B$. Similarly, we set $\Delta_+ = -\Delta_-$ to maintain the opposite parity of the p-wave couplings, just as in the nanowire case. The reason for keeping $\mu_- = 0$ shall be made clear soon. At $\mu_+=0$ both chains are in the topological regime and have two unpaired Majoranas at their ends. However, due to the presence of an s-wave inter-chain coupling $\Delta_s$, the Majoranas at each end pair up to form a normal fermion, rendering the spinfull Kitaev chain into a trivial phase. This can be seen in Fig. \ref{fig:6}(a), where the yellow unpaired Majoranas hybridize through the solid black bonds caused by $\Delta_s$. \\
\indent Figure \ref{fig:6}(b) represents an intermediate stage where $\mu_+$ is finite and can be seen as the green bonds in the top two chains. Fig. \ref{fig:6}(c) depicts the $\mu_+ = \infty$ case. Here, $\mu_+$ acts as the only dominating bond in the $+$ chain and decouples the chain $+$ chain from the $-$ chain. The $+$ chain goes into a trivial regime while the $-$ chain stays in the topological regime, allowing there to be effectively two Majoranas at the ends of the wire. This is what precisely happens in the nanowire system, where the effective onsite potential being controlled by the magnetic field $B$ of one band tunes it towards the topological regimes while tunes the other band towards the trivial regime. Before the transition, there are two p-wave superconductors hosting Majoranas. However, their interaction effectively leads to none at the edges. At the topological transition, one band becomes trivial, leading to two unpaired Majoranas at the ends. \\
\indent Figure \ref{fig:7} shows the entanglement entropy plots for the spinfull Kitaev chain under the settings described above. Figure  \ref{fig:6} (a) shows the BEE plot for a 100-site chain while Fig. \ref{fig:7} (b) shows the DEE plot for the same chain. Comparing Fig. \ref{fig:3} (a) and Fig. \ref{fig:7} (a) we see very similar features for both the spinfull Kitaev chain and the nanowire with no quantization and negligible change in the entanglement entropy value over the phase transition. Similarly, comparing Fig. \ref{fig:3} (b) and Fig. \ref{fig:7} (b) we see that the DEE is zero in the trivial regime and quantized to $ln2$ in the topological regime in both the spinfull Kitaev chain and the nanowire. A DEE value of $ln2$ in the topological regime is logically expected in the spinfull Kitaev chain since chains are now independent of any inter-chain interaction, the $+$ Kitaev chain, being in its trivial regime, contributes $0$ units to the DEE while the $-$ Kitaev chain, being topological, contributes $ln2$ units to the DEE. \\
% \begin{figure}[!tbp]
% 	\centering
%     \includegraphics[width=1.0\textwidth]{9.png}\label{9}

% 	\caption{(a) the DEE (in units of $\ln2$) of a $4 \mu m$ disordered nanowire, and (b) the DEE (in units of $\ln2$) of an $8 \mu m$ disordered nanowire plotted against $B/\Delta$. The peak value and variance for the gaussian smooth potentials have been taken from Table 1. The occurrence if near zero-energy states before the topological phase transition in (a) and (b) can be seen in Figures 8(a) and 8(b) respectively. Nonetheless, the DEE signature remains near zero in the trivial regime and only attains $ln2$ quantization after the true topological phase transition. Hence the DEE remains faithful even in the presence of inhomogeneous smooth potentials at the ends for both short and long nanowires. }
    
% 	\label{fig:9}
% \end{figure}
\indent Next, we focus on the effects of disorder in the nanowire system and its entanglement entropy. As mentioned earlier, the disorder that we study in this work, i.e., smooth inhomogeneous potentials at the ends of the nanowire, causes the appearance of near zero-energy sub-gap states called Andreev bound state (ABS) or quasi-MZMs. These states are trivial in nature and the near zero-energy behavior of these trivial states over wide ranges of $B$ and $\mu$ hinders the conclusive detection of MZMs through present protocols.\\
% \begin{figure}[!tbp]
% 	\centering
%     \includegraphics[width=1.0\textwidth]{10.png}\label{10}

% 	\caption{ The phase diagram of the disconnected entanglement entropy (in units of $\ln 2$) plotted against the magnetic field $B$ and the chemical potential $\mu$ for (a) a $4\mu m$ disordered nanowire and (b) an $8\mu m$ disordered nanowire. The dashed black curve represents the theoretically calculated topological phase boundary. The DEE faithfully detects phase transitions, even in the presence of inhomogeneous smooth potentials in the ends of the nanowire, for a reasonable range of experimentally controllable parameters $\mu$ and $B$.}
    
% 	\label{fig:10}
% \end{figure}
\indent First, in Fig. \ref{fig:8}, we show the energy eigenspectrum of the nanowire in the presence of smooth potentials $V(x)$ at the ends. Figure \ref{fig:8}(a) and Fig. \ref{fig:8}(b) show the energy eigenspectrum for a 4$\mu m$ and an 8$\mu m$ wire, respectively. As shown in in Fig. \ref{fig:8}(a), the system hosts zero-energy modes before the topological phase transition (highlighted by the dashed red line), depicting the presence of disorder-induced trivial zero-energy modes. These states become more prominent and appear earlier as the system length increases, as is visible in the longer nanowire case (Fig. \ref{fig:8}(b)). \\
\indent In Fig. \ref{fig:8}(c) and Fig. \ref{fig:8}(d), we show the DEE of the 4$\mu m$ and an 8$\mu m$ disordered nanowires corresponding to the energy spectra in Figs. \ref{fig:8}(a) and  \ref{fig:8}(b). For both the nanowire lengths, the DEE is zero in the trivial phase and rises to a quantized value of $ln2$ in the topological phase only, similar to the pristine nanowire case. The presence of near-zero energy trivial states has only a very subtle effect on DEE. It causes it to slightly overshoot at the phase transition, which may be related to a very generic peaking behavior of entanglement entropy at quantum critical points in one-dimensional systems. Nevertheless, the DEE remains faithful in signalling MZMs and the topological phase transition with clarity in cases where near-zero energy states are present. We also plot the phase diagram of DEE in the disordered case. Figures \ref{fig:8}(e) and \ref{fig:8}(f) depict the phase diagram of DEE for a 4$\mu m$ and an 8$\mu m$ wire, respectively, and highlight that the transition in DEE exactly matches the theoretically estimated topological phase boundary. Thus, the DEE conclusively detects topological phase transitions, even in the presence of disorder, for a wide range of parameters $B$ and $\mu$.  \\
\indent The disconnected fermion parity noise $\delta P$, a physical observable closely related to the DEE, is a possible candidate to effectively gauge DEE in experiments. Figure \ref{fig:11} depicts the variation of $\delta P$ as a function of the Zeeman field for the nanowire. In Fig. \ref{fig:11}(a) and \ref{fig:11}(b), we plot the disconnected fermion parity noise of a pristine $4 \mu m$ nanowire and a pristine $8 \mu m$ nanowire against the magnetic field. For both nanowire lengths, we observe that the signatures closely match the DEE signature, thus pointing out the connection between DEE and the fermion parity noise. \\
\indent To study the robustness of this physically observable metric to disorder, we show the corresponding plots for the disordered nanowire case in Figs. \ref{fig:11}(c) and \ref{fig:11}(d). As expected, we observe that the signatures stay robust to the presence of trivial zero-energy modes and closely match the DEE signatures. We also plot the phase diagrams of the disconnected fermion parity noise for Fig. \ref{fig:12}(a) a $4\mu m$ pristine nanowire, Fig. \ref{fig:12}(b) an $8\mu m$ pristine nanowire, Fig. \ref{fig:12}(c) a $4\mu m$ disordered nanowire, Fig. \ref{fig:12}(d) an $8\mu m$ disordered nanowire and show that the fermion parity noise signature remains clean and robust over a good range of controllable parameters $\mu$ and $B$ hence acting as a faithful detector of topological phase transitions. \\
\indent Lastly, we find it worthwhile to demonstrate the superiority of entanglement entropy (and hence observables such as the fermion parity) over techniques with current experimental attention, such as local and non-local differential conductance. In Fig. \ref{fig:13}, we depict the local conductance ($G_{LL}$), non-local conductance ($G_{LR}$), and the DEE ($S_D$) of a three-terminal N-TS-N setup involving an intermediate length $6 \mu m$ nanowire with minimal to modest disorder. Even under low disorder, the local conductance fails to detect the phase transition with a premature closure, as shown in Fig. \ref{fig:13}(a). The non-local conductance signatures, on the other hand, become obscure and faint under experimentally relevant measurement precision, unable to pinpoint the critical field, as depicted in Fig. \ref{fig:13}(b). The DEE, however, remains faithful and accurately signals the phase transition, as shown in Fig. \ref{fig:13}(c). 
\section{Conclusion} 
We proposed and analyzed in depth the use of entanglement entropy and its close relation with topological order for a fool-proof signalling of MZMs in Rashba nanowires. We showed that although the BEE is incapable of determining topological phase transitions, owing to volume and area contributions, the DEE reliably pinpoints topological phase transitions following the distillation of the size contributions. The DEE signatures stay robust to system size and controllable parameters such as chemical potential and magnetic field, even in the presence of disorders such as smooth inhomogeneous potentials at the ends of the nanowire relevant to experimental setups. We explained the quantization of the entanglement entropy in a bottom-up fashion by deconstructing Rashba nanowires as interacting Kitaev Chains. Further, we connected the entanglement entropy signatures to a conserved observable of the system, the fermion parity noise, and illustrated that it displays signatures similar to the entanglement entropy. Finally, we concluded by elucidating a simple case with minimal disorder and moderate nanowire lengths where the local conductance signature closes prematurely, and the non-local conductance signature is obscure, but the DEE signature is conclusive and faithfully signals the topological transition. The theoretical methods developed here are of significant relevance toward making experimental progress for the conclusive detection of MZMs, and looking beyond conductance spectroscopy measurements. \\
{\it{Acknowedgements:}}  The authors acknowledge the Visvesvaraya PhD Scheme of the Ministry of Electronics and Information Technology (MEITY), Government of India, as implemented by Digital India Corporation (formerly Media Lab Asia). This work is also supported by the Science and Engineering Research Board (SERB), Government of India, Grant No. Grant No. STR/2019/000030, the Ministry of Human Resource Development (MHRD), Government of India, Grant No. STARS/APR2019/NS/226/FS under the STARS scheme.
\bibliography{main}

\providecommand{\noopsort}[1]{}\providecommand{\singleletter}[1]{#1}%
\begin{thebibliography}{50}%
\makeatletter
\providecommand \@ifxundefined [1]{%
 \@ifx{#1\undefined}
}%
\providecommand \@ifnum [1]{%
 \ifnum #1\expandafter \@firstoftwo
 \else \expandafter \@secondoftwo
 \fi
}%
\providecommand \@ifx [1]{%
 \ifx #1\expandafter \@firstoftwo
 \else \expandafter \@secondoftwo
 \fi
}%
\providecommand \natexlab [1]{#1}%
\providecommand \enquote  [1]{``#1''}%
\providecommand \bibnamefont  [1]{#1}%
\providecommand \bibfnamefont [1]{#1}%
\providecommand \citenamefont [1]{#1}%
\providecommand \href@noop [0]{\@secondoftwo}%
\providecommand \href [0]{\begingroup \@sanitize@url \@href}%
\providecommand \@href[1]{\@@startlink{#1}\@@href}%
\providecommand \@@href[1]{\endgroup#1\@@endlink}%
\providecommand \@sanitize@url [0]{\catcode `\\12\catcode `\$12\catcode
  `\&12\catcode `\#12\catcode `\^12\catcode `\_12\catcode `\%12\relax}%
\providecommand \@@startlink[1]{}%
\providecommand \@@endlink[0]{}%
\providecommand \url  [0]{\begingroup\@sanitize@url \@url }%
\providecommand \@url [1]{\endgroup\@href {#1}{\urlprefix }}%
\providecommand \urlprefix  [0]{URL }%
\providecommand \Eprint [0]{\href }%
\providecommand \doibase [0]{https://doi.org/}%
\providecommand \selectlanguage [0]{\@gobble}%
\providecommand \bibinfo  [0]{\@secondoftwo}%
\providecommand \bibfield  [0]{\@secondoftwo}%
\providecommand \translation [1]{[#1]}%
\providecommand \BibitemOpen [0]{}%
\providecommand \bibitemStop [0]{}%
\providecommand \bibitemNoStop [0]{.\EOS\space}%
\providecommand \EOS [0]{\spacefactor3000\relax}%
\providecommand \BibitemShut  [1]{\csname bibitem#1\endcsname}%
\let\auto@bib@innerbib\@empty
%</preamble>
\bibitem [{\citenamefont {Kitaev}(2001)}]{kitaev:physusp2001}%
  \BibitemOpen
  \bibfield  {author} {\bibinfo {author} {\bibfnamefont {A.~Y.}\ \bibnamefont
  {Kitaev}},\ }\href {http://stacks.iop.org/1063-7869/44/i=10S/a=S29}
  {\bibfield  {journal} {\bibinfo  {journal} {Physics-Uspekhi}\ }\textbf
  {\bibinfo {volume} {44}},\ \bibinfo {pages} {131} (\bibinfo {year}
  {2001})}\BibitemShut {NoStop}%
\bibitem [{\citenamefont {Aasen}\ \emph {et~al.}(2016)\citenamefont {Aasen},
  \citenamefont {Hell}, \citenamefont {Mishmash}, \citenamefont {Higginbotham},
  \citenamefont {Danon}, \citenamefont {Leijnse}, \citenamefont {Jespersen},
  \citenamefont {Folk}, \citenamefont {Marcus}, \citenamefont {Flensberg},\
  and\ \citenamefont {Alicea}}]{Aasen-2016}%
  \BibitemOpen
  \bibfield  {author} {\bibinfo {author} {\bibfnamefont {D.}~\bibnamefont
  {Aasen}}, \bibinfo {author} {\bibfnamefont {M.}~\bibnamefont {Hell}},
  \bibinfo {author} {\bibfnamefont {R.~V.}\ \bibnamefont {Mishmash}}, \bibinfo
  {author} {\bibfnamefont {A.}~\bibnamefont {Higginbotham}}, \bibinfo {author}
  {\bibfnamefont {J.}~\bibnamefont {Danon}}, \bibinfo {author} {\bibfnamefont
  {M.}~\bibnamefont {Leijnse}}, \bibinfo {author} {\bibfnamefont {T.~S.}\
  \bibnamefont {Jespersen}}, \bibinfo {author} {\bibfnamefont {J.~A.}\
  \bibnamefont {Folk}}, \bibinfo {author} {\bibfnamefont {C.~M.}\ \bibnamefont
  {Marcus}}, \bibinfo {author} {\bibfnamefont {K.}~\bibnamefont {Flensberg}},\
  and\ \bibinfo {author} {\bibfnamefont {J.}~\bibnamefont {Alicea}},\ }\href
  {https://doi.org/10.1103/PhysRevX.6.031016} {\bibfield  {journal} {\bibinfo
  {journal} {Phys. Rev. X}\ }\textbf {\bibinfo {volume} {6}},\ \bibinfo {pages}
  {031016} (\bibinfo {year} {2016})}\BibitemShut {NoStop}%
\bibitem [{\citenamefont {O{\rq}Brien}\ \emph {et~al.}(2018)\citenamefont
  {O{\rq}Brien}, \citenamefont {Rożek},\ and\ \citenamefont
  {Akhmerov}}]{obrien:prl2018}%
  \BibitemOpen
  \bibfield  {author} {\bibinfo {author} {\bibfnamefont {T.~E.}\ \bibnamefont
  {O{\rq}Brien}}, \bibinfo {author} {\bibfnamefont {P.}~\bibnamefont
  {Rożek}},\ and\ \bibinfo {author} {\bibfnamefont {A.~R.}\ \bibnamefont
  {Akhmerov}},\ }\href
  {https://doi.org/https://link.aps.org/doi/10.1103/PhysRevLett.120.220504}
  {\bibfield  {journal} {\bibinfo  {journal} {Phys. Rev. Lett.}\ }\textbf
  {\bibinfo {volume} {120}},\ \bibinfo {pages} {220504} (\bibinfo {year}
  {2018})}\BibitemShut {NoStop}%
\bibitem [{\citenamefont {Nayak}\ \emph {et~al.}(2008)\citenamefont {Nayak},
  \citenamefont {Simon}, \citenamefont {Stern}, \citenamefont {Freedman},\ and\
  \citenamefont {Das~Sarma}}]{RevModPhys.80.1083}%
  \BibitemOpen
  \bibfield  {author} {\bibinfo {author} {\bibfnamefont {C.}~\bibnamefont
  {Nayak}}, \bibinfo {author} {\bibfnamefont {S.~H.}\ \bibnamefont {Simon}},
  \bibinfo {author} {\bibfnamefont {A.}~\bibnamefont {Stern}}, \bibinfo
  {author} {\bibfnamefont {M.}~\bibnamefont {Freedman}},\ and\ \bibinfo
  {author} {\bibfnamefont {S.}~\bibnamefont {Das~Sarma}},\ }\href
  {https://doi.org/10.1103/RevModPhys.80.1083} {\bibfield  {journal} {\bibinfo
  {journal} {Rev. Mod. Phys.}\ }\textbf {\bibinfo {volume} {80}},\ \bibinfo
  {pages} {1083} (\bibinfo {year} {2008})}\BibitemShut {NoStop}%
\bibitem [{\citenamefont {Sarma}\ \emph {et~al.}(2015)\citenamefont {Sarma},
  \citenamefont {Freedman},\ and\ \citenamefont {Nayak}}]{sarma2015majorana}%
  \BibitemOpen
  \bibfield  {author} {\bibinfo {author} {\bibfnamefont {S.~D.}\ \bibnamefont
  {Sarma}}, \bibinfo {author} {\bibfnamefont {M.}~\bibnamefont {Freedman}},\
  and\ \bibinfo {author} {\bibfnamefont {C.}~\bibnamefont {Nayak}},\
  }\href@noop {} {\bibfield  {journal} {\bibinfo  {journal} {npj Quantum
  Information}\ }\textbf {\bibinfo {volume} {1}},\ \bibinfo {pages} {1}
  (\bibinfo {year} {2015})}\BibitemShut {NoStop}%
\bibitem [{\citenamefont {Marra}(2022)}]{doi:10.1063/5.0102999}%
  \BibitemOpen
  \bibfield  {author} {\bibinfo {author} {\bibfnamefont {P.}~\bibnamefont
  {Marra}},\ }\href {https://doi.org/10.1063/5.0102999} {\bibfield  {journal}
  {\bibinfo  {journal} {Journal of Applied Physics}\ }\textbf {\bibinfo
  {volume} {132}},\ \bibinfo {pages} {231101} (\bibinfo {year} {2022})},\
  \Eprint {https://arxiv.org/abs/https://doi.org/10.1063/5.0102999}
  {https://doi.org/10.1063/5.0102999} \BibitemShut {NoStop}%
\bibitem [{\citenamefont {Das}\ \emph {et~al.}(2012)\citenamefont {Das},
  \citenamefont {Ronen}, \citenamefont {Most}, \citenamefont {Oreg},
  \citenamefont {Heiblum},\ and\ \citenamefont {Shtrikman}}]{Das2012}%
  \BibitemOpen
  \bibfield  {author} {\bibinfo {author} {\bibfnamefont {A.}~\bibnamefont
  {Das}}, \bibinfo {author} {\bibfnamefont {Y.}~\bibnamefont {Ronen}}, \bibinfo
  {author} {\bibfnamefont {Y.}~\bibnamefont {Most}}, \bibinfo {author}
  {\bibfnamefont {Y.}~\bibnamefont {Oreg}}, \bibinfo {author} {\bibfnamefont
  {M.}~\bibnamefont {Heiblum}},\ and\ \bibinfo {author} {\bibfnamefont
  {H.}~\bibnamefont {Shtrikman}},\ }\href {https://doi.org/10.1038/nphys2479}
  {\bibfield  {journal} {\bibinfo  {journal} {Nature Physics}\ }\textbf
  {\bibinfo {volume} {8}},\ \bibinfo {pages} {887} (\bibinfo {year}
  {2012})}\BibitemShut {NoStop}%
\bibitem [{\citenamefont {Mourik}\ \emph {et~al.}(2012)\citenamefont {Mourik},
  \citenamefont {Zuo}, \citenamefont {Frolov}, \citenamefont {Plissard},
  \citenamefont {Bakkers},\ and\ \citenamefont {Kouwenhoven}}]{Mourik-2012}%
  \BibitemOpen
  \bibfield  {author} {\bibinfo {author} {\bibfnamefont {V.}~\bibnamefont
  {Mourik}}, \bibinfo {author} {\bibfnamefont {K.}~\bibnamefont {Zuo}},
  \bibinfo {author} {\bibfnamefont {S.}~\bibnamefont {Frolov}}, \bibinfo
  {author} {\bibfnamefont {S.}~\bibnamefont {Plissard}}, \bibinfo {author}
  {\bibfnamefont {E.}~\bibnamefont {Bakkers}},\ and\ \bibinfo {author}
  {\bibfnamefont {L.}~\bibnamefont {Kouwenhoven}},\ }\href
  {https://doi.org/10.1126/science.1222360} {\bibfield  {journal} {\bibinfo
  {journal} {Science}\ }\textbf {\bibinfo {volume} {336}},\ \bibinfo {pages}
  {1003} (\bibinfo {year} {2012})}\BibitemShut {NoStop}%
\bibitem [{\citenamefont {Deng}\ \emph
  {et~al.}(2016{\natexlab{a}})\citenamefont {Deng}, \citenamefont {S.},
  \citenamefont {Hansen}, \citenamefont {Danon}, \citenamefont {Leijnse},
  \citenamefont {Flensberg}, \citenamefont {Nyg\r{a}rd}, \citenamefont
  {Krogstrup},\ and\ \citenamefont {Marcus}}]{Deng-2016}%
  \BibitemOpen
  \bibfield  {author} {\bibinfo {author} {\bibfnamefont {M.~T.}\ \bibnamefont
  {Deng}}, \bibinfo {author} {\bibfnamefont {V.}~\bibnamefont {S.}}, \bibinfo
  {author} {\bibfnamefont {E.}~\bibnamefont {Hansen}}, \bibinfo {author}
  {\bibfnamefont {J.}~\bibnamefont {Danon}}, \bibinfo {author} {\bibfnamefont
  {M.}~\bibnamefont {Leijnse}}, \bibinfo {author} {\bibfnamefont
  {K.}~\bibnamefont {Flensberg}}, \bibinfo {author} {\bibfnamefont
  {J.}~\bibnamefont {Nyg\r{a}rd}}, \bibinfo {author} {\bibfnamefont
  {P.}~\bibnamefont {Krogstrup}},\ and\ \bibinfo {author} {\bibfnamefont
  {C.~M.}\ \bibnamefont {Marcus}},\ }\href {doi.org/10.1126/science.aaf3961}
  {\bibfield  {journal} {\bibinfo  {journal} {Science}\ }\textbf {\bibinfo
  {volume} {354}},\ \bibinfo {pages} {1557} (\bibinfo {year}
  {2016}{\natexlab{a}})}\BibitemShut {NoStop}%
\bibitem [{\citenamefont {Finck}\ \emph {et~al.}(2013)\citenamefont {Finck},
  \citenamefont {Van~Harlingen}, \citenamefont {Mohseni}, \citenamefont
  {Jung},\ and\ \citenamefont {Li}}]{PhysRevLett.110.126406}%
  \BibitemOpen
  \bibfield  {author} {\bibinfo {author} {\bibfnamefont {A.~D.~K.}\
  \bibnamefont {Finck}}, \bibinfo {author} {\bibfnamefont {D.~J.}\ \bibnamefont
  {Van~Harlingen}}, \bibinfo {author} {\bibfnamefont {P.~K.}\ \bibnamefont
  {Mohseni}}, \bibinfo {author} {\bibfnamefont {K.}~\bibnamefont {Jung}},\ and\
  \bibinfo {author} {\bibfnamefont {X.}~\bibnamefont {Li}},\ }\href
  {https://doi.org/10.1103/PhysRevLett.110.126406} {\bibfield  {journal}
  {\bibinfo  {journal} {Phys. Rev. Lett.}\ }\textbf {\bibinfo {volume} {110}},\
  \bibinfo {pages} {126406} (\bibinfo {year} {2013})}\BibitemShut {NoStop}%
\bibitem [{\citenamefont {Nichele}\ \emph {et~al.}(2017)\citenamefont
  {Nichele}, \citenamefont {Drachmann}, \citenamefont {Whiticar}, \citenamefont
  {O'Farrell}, \citenamefont {Suominen}, \citenamefont {Fornieri},
  \citenamefont {Wang}, \citenamefont {Gardner}, \citenamefont {Thomas},
  \citenamefont {Hatke}, \citenamefont {Krogstrup}, \citenamefont {Manfra},
  \citenamefont {Flensberg},\ and\ \citenamefont
  {Marcus}}]{Scaling_ZBP_Marcus}%
  \BibitemOpen
  \bibfield  {author} {\bibinfo {author} {\bibfnamefont {F.}~\bibnamefont
  {Nichele}}, \bibinfo {author} {\bibfnamefont {A.~C.~C.}\ \bibnamefont
  {Drachmann}}, \bibinfo {author} {\bibfnamefont {A.~M.}\ \bibnamefont
  {Whiticar}}, \bibinfo {author} {\bibfnamefont {E.~C.~T.}\ \bibnamefont
  {O'Farrell}}, \bibinfo {author} {\bibfnamefont {H.~J.}\ \bibnamefont
  {Suominen}}, \bibinfo {author} {\bibfnamefont {A.}~\bibnamefont {Fornieri}},
  \bibinfo {author} {\bibfnamefont {T.}~\bibnamefont {Wang}}, \bibinfo {author}
  {\bibfnamefont {G.~C.}\ \bibnamefont {Gardner}}, \bibinfo {author}
  {\bibfnamefont {C.}~\bibnamefont {Thomas}}, \bibinfo {author} {\bibfnamefont
  {A.~T.}\ \bibnamefont {Hatke}}, \bibinfo {author} {\bibfnamefont
  {P.}~\bibnamefont {Krogstrup}}, \bibinfo {author} {\bibfnamefont {M.~J.}\
  \bibnamefont {Manfra}}, \bibinfo {author} {\bibfnamefont {K.}~\bibnamefont
  {Flensberg}},\ and\ \bibinfo {author} {\bibfnamefont {C.~M.}\ \bibnamefont
  {Marcus}},\ }\href {https://doi.org/10.1103/PhysRevLett.119.136803}
  {\bibfield  {journal} {\bibinfo  {journal} {Phys. Rev. Lett.}\ }\textbf
  {\bibinfo {volume} {119}},\ \bibinfo {pages} {136803} (\bibinfo {year}
  {2017})}\BibitemShut {NoStop}%
\bibitem [{\citenamefont {Albrecht}\ \emph {et~al.}(2016)\citenamefont
  {Albrecht}, \citenamefont {Higginbotham}, \citenamefont {Madsen},
  \citenamefont {Kuemmeth}, \citenamefont {Jespersen}, \citenamefont
  {Nyg{\aa}rd}, \citenamefont {Krogstrup},\ and\ \citenamefont
  {Marcus}}]{Albrecht2016}%
  \BibitemOpen
  \bibfield  {author} {\bibinfo {author} {\bibfnamefont {S.~M.}\ \bibnamefont
  {Albrecht}}, \bibinfo {author} {\bibfnamefont {A.~P.}\ \bibnamefont
  {Higginbotham}}, \bibinfo {author} {\bibfnamefont {M.}~\bibnamefont
  {Madsen}}, \bibinfo {author} {\bibfnamefont {F.}~\bibnamefont {Kuemmeth}},
  \bibinfo {author} {\bibfnamefont {T.~S.}\ \bibnamefont {Jespersen}}, \bibinfo
  {author} {\bibfnamefont {J.}~\bibnamefont {Nyg{\aa}rd}}, \bibinfo {author}
  {\bibfnamefont {P.}~\bibnamefont {Krogstrup}},\ and\ \bibinfo {author}
  {\bibfnamefont {C.~M.}\ \bibnamefont {Marcus}},\ }\href
  {https://doi.org/10.1038/nature17162} {\bibfield  {journal} {\bibinfo
  {journal} {Nature}\ }\textbf {\bibinfo {volume} {531}},\ \bibinfo {pages}
  {206} (\bibinfo {year} {2016})}\BibitemShut {NoStop}%
\bibitem [{\citenamefont {Vuik}\ \emph {et~al.}(2019)\citenamefont {Vuik},
  \citenamefont {Nijholt}, \citenamefont {Akhmerov},\ and\ \citenamefont
  {Wimmer}}]{10.21468/SciPostPhys.7.5.061}%
  \BibitemOpen
  \bibfield  {author} {\bibinfo {author} {\bibfnamefont {A.}~\bibnamefont
  {Vuik}}, \bibinfo {author} {\bibfnamefont {B.}~\bibnamefont {Nijholt}},
  \bibinfo {author} {\bibfnamefont {A.~R.}\ \bibnamefont {Akhmerov}},\ and\
  \bibinfo {author} {\bibfnamefont {M.}~\bibnamefont {Wimmer}},\ }\href
  {https://doi.org/10.21468/SciPostPhys.7.5.061} {\bibfield  {journal}
  {\bibinfo  {journal} {SciPost Phys.}\ }\textbf {\bibinfo {volume} {7}},\
  \bibinfo {pages} {61} (\bibinfo {year} {2019})}\BibitemShut {NoStop}%
\bibitem [{\citenamefont {Kells}\ \emph {et~al.}(2012)\citenamefont {Kells},
  \citenamefont {Meidan},\ and\ \citenamefont {Brouwer}}]{PhysRevB.86.100503}%
  \BibitemOpen
  \bibfield  {author} {\bibinfo {author} {\bibfnamefont {G.}~\bibnamefont
  {Kells}}, \bibinfo {author} {\bibfnamefont {D.}~\bibnamefont {Meidan}},\ and\
  \bibinfo {author} {\bibfnamefont {P.~W.}\ \bibnamefont {Brouwer}},\ }\href
  {https://doi.org/10.1103/PhysRevB.86.100503} {\bibfield  {journal} {\bibinfo
  {journal} {Phys. Rev. B}\ }\textbf {\bibinfo {volume} {86}},\ \bibinfo
  {pages} {100503} (\bibinfo {year} {2012})}\BibitemShut {NoStop}%
\bibitem [{\citenamefont {Fleckenstein}\ \emph {et~al.}(2018)\citenamefont
  {Fleckenstein}, \citenamefont {Dom\'{\i}nguez}, \citenamefont
  {Traverso~Ziani},\ and\ \citenamefont {Trauzettel}}]{PhysRevB.97.155425}%
  \BibitemOpen
  \bibfield  {author} {\bibinfo {author} {\bibfnamefont {C.}~\bibnamefont
  {Fleckenstein}}, \bibinfo {author} {\bibfnamefont {F.}~\bibnamefont
  {Dom\'{\i}nguez}}, \bibinfo {author} {\bibfnamefont {N.}~\bibnamefont
  {Traverso~Ziani}},\ and\ \bibinfo {author} {\bibfnamefont {B.}~\bibnamefont
  {Trauzettel}},\ }\href {https://doi.org/10.1103/PhysRevB.97.155425}
  {\bibfield  {journal} {\bibinfo  {journal} {Phys. Rev. B}\ }\textbf {\bibinfo
  {volume} {97}},\ \bibinfo {pages} {155425} (\bibinfo {year}
  {2018})}\BibitemShut {NoStop}%
\bibitem [{\citenamefont {Liu}\ \emph {et~al.}(2017)\citenamefont {Liu},
  \citenamefont {Sau}, \citenamefont {Stanescu},\ and\ \citenamefont
  {Das~Sarma}}]{PhysRevB.96.075161}%
  \BibitemOpen
  \bibfield  {author} {\bibinfo {author} {\bibfnamefont {C.-X.}\ \bibnamefont
  {Liu}}, \bibinfo {author} {\bibfnamefont {J.~D.}\ \bibnamefont {Sau}},
  \bibinfo {author} {\bibfnamefont {T.~D.}\ \bibnamefont {Stanescu}},\ and\
  \bibinfo {author} {\bibfnamefont {S.}~\bibnamefont {Das~Sarma}},\ }\href
  {https://doi.org/10.1103/PhysRevB.96.075161} {\bibfield  {journal} {\bibinfo
  {journal} {Phys. Rev. B}\ }\textbf {\bibinfo {volume} {96}},\ \bibinfo
  {pages} {075161} (\bibinfo {year} {2017})}\BibitemShut {NoStop}%
\bibitem [{\citenamefont {Lobos}\ and\ \citenamefont {Sarma}(2015)}]{Lobos}%
  \BibitemOpen
  \bibfield  {author} {\bibinfo {author} {\bibfnamefont {A.~M.}\ \bibnamefont
  {Lobos}}\ and\ \bibinfo {author} {\bibfnamefont {S.~D.}\ \bibnamefont
  {Sarma}},\ }\href@noop {} {\bibfield  {journal} {\bibinfo  {journal} {New
  Journal of Physics}\ }\textbf {\bibinfo {volume} {17}},\ \bibinfo {pages}
  {065010} (\bibinfo {year} {2015})}\BibitemShut {NoStop}%
\bibitem [{\citenamefont {Cayao}\ \emph {et~al.}(2015)\citenamefont {Cayao},
  \citenamefont {Prada}, \citenamefont {San-Jose},\ and\ \citenamefont
  {Aguado}}]{PhysRevB.91.024514}%
  \BibitemOpen
  \bibfield  {author} {\bibinfo {author} {\bibfnamefont {J.}~\bibnamefont
  {Cayao}}, \bibinfo {author} {\bibfnamefont {E.}~\bibnamefont {Prada}},
  \bibinfo {author} {\bibfnamefont {P.}~\bibnamefont {San-Jose}},\ and\
  \bibinfo {author} {\bibfnamefont {R.}~\bibnamefont {Aguado}},\ }\href
  {https://doi.org/10.1103/PhysRevB.91.024514} {\bibfield  {journal} {\bibinfo
  {journal} {Phys. Rev. B}\ }\textbf {\bibinfo {volume} {91}},\ \bibinfo
  {pages} {024514} (\bibinfo {year} {2015})}\BibitemShut {NoStop}%
\bibitem [{\citenamefont {San-Jose}\ \emph {et~al.}(2016)\citenamefont
  {San-Jose}, \citenamefont {Cayao}, \citenamefont {Prada},\ and\ \citenamefont
  {Aguado}}]{San-Jose2016}%
  \BibitemOpen
  \bibfield  {author} {\bibinfo {author} {\bibfnamefont {P.}~\bibnamefont
  {San-Jose}}, \bibinfo {author} {\bibfnamefont {J.}~\bibnamefont {Cayao}},
  \bibinfo {author} {\bibfnamefont {E.}~\bibnamefont {Prada}},\ and\ \bibinfo
  {author} {\bibfnamefont {R.}~\bibnamefont {Aguado}},\ }\href
  {https://doi.org/10.1038/srep21427} {\bibfield  {journal} {\bibinfo
  {journal} {Scientific Reports}\ }\textbf {\bibinfo {volume} {6}},\ \bibinfo
  {pages} {21427} (\bibinfo {year} {2016})}\BibitemShut {NoStop}%
\bibitem [{\citenamefont {Hess}\ \emph {et~al.}(2021)\citenamefont {Hess},
  \citenamefont {Legg}, \citenamefont {Loss},\ and\ \citenamefont
  {Klinovaja}}]{PhysRevB.104.075405}%
  \BibitemOpen
  \bibfield  {author} {\bibinfo {author} {\bibfnamefont {R.}~\bibnamefont
  {Hess}}, \bibinfo {author} {\bibfnamefont {H.~F.}\ \bibnamefont {Legg}},
  \bibinfo {author} {\bibfnamefont {D.}~\bibnamefont {Loss}},\ and\ \bibinfo
  {author} {\bibfnamefont {J.}~\bibnamefont {Klinovaja}},\ }\href
  {https://doi.org/10.1103/PhysRevB.104.075405} {\bibfield  {journal} {\bibinfo
   {journal} {Phys. Rev. B}\ }\textbf {\bibinfo {volume} {104}},\ \bibinfo
  {pages} {075405} (\bibinfo {year} {2021})}\BibitemShut {NoStop}%
\bibitem [{\citenamefont {Pan}\ \emph {et~al.}(2021{\natexlab{a}})\citenamefont
  {Pan}, \citenamefont {Liu}, \citenamefont {Wimmer},\ and\ \citenamefont
  {Das~Sarma}}]{PhysRevB.103.214502}%
  \BibitemOpen
  \bibfield  {author} {\bibinfo {author} {\bibfnamefont {H.}~\bibnamefont
  {Pan}}, \bibinfo {author} {\bibfnamefont {C.-X.}\ \bibnamefont {Liu}},
  \bibinfo {author} {\bibfnamefont {M.}~\bibnamefont {Wimmer}},\ and\ \bibinfo
  {author} {\bibfnamefont {S.}~\bibnamefont {Das~Sarma}},\ }\href
  {https://doi.org/10.1103/PhysRevB.103.214502} {\bibfield  {journal} {\bibinfo
   {journal} {Phys. Rev. B}\ }\textbf {\bibinfo {volume} {103}},\ \bibinfo
  {pages} {214502} (\bibinfo {year} {2021}{\natexlab{a}})}\BibitemShut
  {NoStop}%
\bibitem [{\citenamefont {Gramich}\ \emph {et~al.}(2017)\citenamefont
  {Gramich}, \citenamefont {Baumgartner},\ and\ \citenamefont
  {Sch\"onenberger}}]{PhysRevB.96.195418}%
  \BibitemOpen
  \bibfield  {author} {\bibinfo {author} {\bibfnamefont {J.}~\bibnamefont
  {Gramich}}, \bibinfo {author} {\bibfnamefont {A.}~\bibnamefont
  {Baumgartner}},\ and\ \bibinfo {author} {\bibfnamefont {C.}~\bibnamefont
  {Sch\"onenberger}},\ }\href {https://doi.org/10.1103/PhysRevB.96.195418}
  {\bibfield  {journal} {\bibinfo  {journal} {Phys. Rev. B}\ }\textbf {\bibinfo
  {volume} {96}},\ \bibinfo {pages} {195418} (\bibinfo {year}
  {2017})}\BibitemShut {NoStop}%
\bibitem [{\citenamefont {Fregoso}\ \emph {et~al.}(2013)\citenamefont
  {Fregoso}, \citenamefont {Lobos},\ and\ \citenamefont
  {Das~Sarma}}]{PhysRevB.88.180507}%
  \BibitemOpen
  \bibfield  {author} {\bibinfo {author} {\bibfnamefont {B.~M.}\ \bibnamefont
  {Fregoso}}, \bibinfo {author} {\bibfnamefont {A.~M.}\ \bibnamefont {Lobos}},\
  and\ \bibinfo {author} {\bibfnamefont {S.}~\bibnamefont {Das~Sarma}},\ }\href
  {https://doi.org/10.1103/PhysRevB.88.180507} {\bibfield  {journal} {\bibinfo
  {journal} {Phys. Rev. B}\ }\textbf {\bibinfo {volume} {88}},\ \bibinfo
  {pages} {180507} (\bibinfo {year} {2013})}\BibitemShut {NoStop}%
\bibitem [{\citenamefont {Rosdahl}\ \emph {et~al.}(2018)\citenamefont
  {Rosdahl}, \citenamefont {Vuik}, \citenamefont {Kjaergaard},\ and\
  \citenamefont {Akhmerov}}]{PhysRevB.97.045421}%
  \BibitemOpen
  \bibfield  {author} {\bibinfo {author} {\bibfnamefont {T.~O.}\ \bibnamefont
  {Rosdahl}}, \bibinfo {author} {\bibfnamefont {A.}~\bibnamefont {Vuik}},
  \bibinfo {author} {\bibfnamefont {M.}~\bibnamefont {Kjaergaard}},\ and\
  \bibinfo {author} {\bibfnamefont {A.~R.}\ \bibnamefont {Akhmerov}},\ }\href
  {https://doi.org/10.1103/PhysRevB.97.045421} {\bibfield  {journal} {\bibinfo
  {journal} {Phys. Rev. B}\ }\textbf {\bibinfo {volume} {97}},\ \bibinfo
  {pages} {045421} (\bibinfo {year} {2018})}\BibitemShut {NoStop}%
\bibitem [{\citenamefont {Puglia}\ \emph {et~al.}(2021)\citenamefont {Puglia},
  \citenamefont {Martinez}, \citenamefont {M\'enard}, \citenamefont {P\"oschl},
  \citenamefont {Gronin}, \citenamefont {Gardner}, \citenamefont {Kallaher},
  \citenamefont {Manfra}, \citenamefont {Marcus}, \citenamefont
  {Higginbotham},\ and\ \citenamefont {Casparis}}]{puglia}%
  \BibitemOpen
  \bibfield  {author} {\bibinfo {author} {\bibfnamefont {D.}~\bibnamefont
  {Puglia}}, \bibinfo {author} {\bibfnamefont {E.~A.}\ \bibnamefont
  {Martinez}}, \bibinfo {author} {\bibfnamefont {G.~C.}\ \bibnamefont
  {M\'enard}}, \bibinfo {author} {\bibfnamefont {A.}~\bibnamefont {P\"oschl}},
  \bibinfo {author} {\bibfnamefont {S.}~\bibnamefont {Gronin}}, \bibinfo
  {author} {\bibfnamefont {G.~C.}\ \bibnamefont {Gardner}}, \bibinfo {author}
  {\bibfnamefont {R.}~\bibnamefont {Kallaher}}, \bibinfo {author}
  {\bibfnamefont {M.~J.}\ \bibnamefont {Manfra}}, \bibinfo {author}
  {\bibfnamefont {C.~M.}\ \bibnamefont {Marcus}}, \bibinfo {author}
  {\bibfnamefont {A.~P.}\ \bibnamefont {Higginbotham}},\ and\ \bibinfo {author}
  {\bibfnamefont {L.}~\bibnamefont {Casparis}},\ }\href
  {https://doi.org/10.1103/PhysRevB.103.235201} {\bibfield  {journal} {\bibinfo
   {journal} {Phys. Rev. B}\ }\textbf {\bibinfo {volume} {103}},\ \bibinfo
  {pages} {235201} (\bibinfo {year} {2021})}\BibitemShut {NoStop}%
\bibitem [{\citenamefont {M\'enard}\ \emph {et~al.}(2020)\citenamefont
  {M\'enard}, \citenamefont {Anselmetti}, \citenamefont {Martinez},
  \citenamefont {Puglia}, \citenamefont {Malinowski}, \citenamefont {Lee},
  \citenamefont {Choi}, \citenamefont {Pendharkar}, \citenamefont
  {Palmstr\o{}m}, \citenamefont {Flensberg}, \citenamefont {Marcus},
  \citenamefont {Casparis},\ and\ \citenamefont
  {Higginbotham}}]{Puglia_Cond_Matrix}%
  \BibitemOpen
  \bibfield  {author} {\bibinfo {author} {\bibfnamefont {G.~C.}\ \bibnamefont
  {M\'enard}}, \bibinfo {author} {\bibfnamefont {G.~L.~R.}\ \bibnamefont
  {Anselmetti}}, \bibinfo {author} {\bibfnamefont {E.~A.}\ \bibnamefont
  {Martinez}}, \bibinfo {author} {\bibfnamefont {D.}~\bibnamefont {Puglia}},
  \bibinfo {author} {\bibfnamefont {F.~K.}\ \bibnamefont {Malinowski}},
  \bibinfo {author} {\bibfnamefont {J.~S.}\ \bibnamefont {Lee}}, \bibinfo
  {author} {\bibfnamefont {S.}~\bibnamefont {Choi}}, \bibinfo {author}
  {\bibfnamefont {M.}~\bibnamefont {Pendharkar}}, \bibinfo {author}
  {\bibfnamefont {C.~J.}\ \bibnamefont {Palmstr\o{}m}}, \bibinfo {author}
  {\bibfnamefont {K.}~\bibnamefont {Flensberg}}, \bibinfo {author}
  {\bibfnamefont {C.~M.}\ \bibnamefont {Marcus}}, \bibinfo {author}
  {\bibfnamefont {L.}~\bibnamefont {Casparis}},\ and\ \bibinfo {author}
  {\bibfnamefont {A.~P.}\ \bibnamefont {Higginbotham}},\ }\href
  {https://doi.org/10.1103/PhysRevLett.124.036802} {\bibfield  {journal}
  {\bibinfo  {journal} {Phys. Rev. Lett.}\ }\textbf {\bibinfo {volume} {124}},\
  \bibinfo {pages} {036802} (\bibinfo {year} {2020})}\BibitemShut {NoStop}%
\bibitem [{\citenamefont {Danon}\ \emph {et~al.}(2020)\citenamefont {Danon},
  \citenamefont {Hellenes}, \citenamefont {Hansen}, \citenamefont {Casparis},
  \citenamefont {Higginbotham},\ and\ \citenamefont
  {Flensberg}}]{Flensberg_Nonlocal}%
  \BibitemOpen
  \bibfield  {author} {\bibinfo {author} {\bibfnamefont {J.}~\bibnamefont
  {Danon}}, \bibinfo {author} {\bibfnamefont {A.~B.}\ \bibnamefont {Hellenes}},
  \bibinfo {author} {\bibfnamefont {E.~B.}\ \bibnamefont {Hansen}}, \bibinfo
  {author} {\bibfnamefont {L.}~\bibnamefont {Casparis}}, \bibinfo {author}
  {\bibfnamefont {A.~P.}\ \bibnamefont {Higginbotham}},\ and\ \bibinfo {author}
  {\bibfnamefont {K.}~\bibnamefont {Flensberg}},\ }\href
  {https://doi.org/10.1103/PhysRevLett.124.036801} {\bibfield  {journal}
  {\bibinfo  {journal} {Phys. Rev. Lett.}\ }\textbf {\bibinfo {volume} {124}},\
  \bibinfo {pages} {036801} (\bibinfo {year} {2020})}\BibitemShut {NoStop}%
\bibitem [{\citenamefont {Pan}\ \emph {et~al.}(2021{\natexlab{b}})\citenamefont
  {Pan}, \citenamefont {Sau},\ and\ \citenamefont
  {Das~Sarma}}]{PhysRevB.103.014513}%
  \BibitemOpen
  \bibfield  {author} {\bibinfo {author} {\bibfnamefont {H.}~\bibnamefont
  {Pan}}, \bibinfo {author} {\bibfnamefont {J.~D.}\ \bibnamefont {Sau}},\ and\
  \bibinfo {author} {\bibfnamefont {S.}~\bibnamefont {Das~Sarma}},\ }\href
  {https://doi.org/10.1103/PhysRevB.103.014513} {\bibfield  {journal} {\bibinfo
   {journal} {Phys. Rev. B}\ }\textbf {\bibinfo {volume} {103}},\ \bibinfo
  {pages} {014513} (\bibinfo {year} {2021}{\natexlab{b}})}\BibitemShut
  {NoStop}%
\bibitem [{\citenamefont {Pikulin}\ \emph {et~al.}(2021)\citenamefont
  {Pikulin}, \citenamefont {van Heck}, \citenamefont {Karzig}, \citenamefont
  {Martinez}, \citenamefont {Nijholt}, \citenamefont {Laeven}, \citenamefont
  {Winkler}, \citenamefont {Watson}, \citenamefont {Heedt}, \citenamefont
  {Temurhan}, \citenamefont {Svidenko}, \citenamefont {Lutchyn}, \citenamefont
  {Thomas}, \citenamefont {de~Lange}, \citenamefont {Casparis},\ and\
  \citenamefont {Nayak}}]{https:MStgp}%
  \BibitemOpen
  \bibfield  {author} {\bibinfo {author} {\bibfnamefont {D.~I.}\ \bibnamefont
  {Pikulin}}, \bibinfo {author} {\bibfnamefont {B.}~\bibnamefont {van Heck}},
  \bibinfo {author} {\bibfnamefont {T.}~\bibnamefont {Karzig}}, \bibinfo
  {author} {\bibfnamefont {E.~A.}\ \bibnamefont {Martinez}}, \bibinfo {author}
  {\bibfnamefont {B.}~\bibnamefont {Nijholt}}, \bibinfo {author} {\bibfnamefont
  {T.}~\bibnamefont {Laeven}}, \bibinfo {author} {\bibfnamefont {G.~W.}\
  \bibnamefont {Winkler}}, \bibinfo {author} {\bibfnamefont {J.~D.}\
  \bibnamefont {Watson}}, \bibinfo {author} {\bibfnamefont {S.}~\bibnamefont
  {Heedt}}, \bibinfo {author} {\bibfnamefont {M.}~\bibnamefont {Temurhan}},
  \bibinfo {author} {\bibfnamefont {V.}~\bibnamefont {Svidenko}}, \bibinfo
  {author} {\bibfnamefont {R.~M.}\ \bibnamefont {Lutchyn}}, \bibinfo {author}
  {\bibfnamefont {M.}~\bibnamefont {Thomas}}, \bibinfo {author} {\bibfnamefont
  {G.}~\bibnamefont {de~Lange}}, \bibinfo {author} {\bibfnamefont
  {L.}~\bibnamefont {Casparis}},\ and\ \bibinfo {author} {\bibfnamefont
  {C.}~\bibnamefont {Nayak}},\ }\href
  {https://doi.org/10.48550/ARXIV.2103.12217} {\bibinfo {title} {Protocol to
  identify a topological superconducting phase in a three-terminal device}}
  (\bibinfo {year} {2021})\BibitemShut {NoStop}%
\bibitem [{\citenamefont {Aghaee}\ \emph {et~al.}(2022)\citenamefont {Aghaee},
  \citenamefont {Akkala}, \citenamefont {Alam}, \citenamefont {Ali},
  \citenamefont {Ramirez}, \citenamefont {Andrzejczuk}, \citenamefont
  {Antipov}, \citenamefont {Aseev}, \citenamefont {Astafev}, \citenamefont
  {Bauer}, \citenamefont {Becker}, \citenamefont {Boddapati}, \citenamefont
  {Boekhout}, \citenamefont {Bommer}, \citenamefont {Hansen}, \citenamefont
  {Bosma}, \citenamefont {Bourdet}, \citenamefont {Boutin}, \citenamefont
  {Caroff}, \citenamefont {Casparis}, \citenamefont {Cassidy}, \citenamefont
  {Christensen}, \citenamefont {Clay}, \citenamefont {Cole}, \citenamefont
  {Corsetti}, \citenamefont {Cui}, \citenamefont {Dalampiras}, \citenamefont
  {Dokania}, \citenamefont {de~Lange}, \citenamefont {de~Moor}, \citenamefont
  {Saldaña}, \citenamefont {Fallahi}, \citenamefont {Fathabad}, \citenamefont
  {Gamble}, \citenamefont {Gardner}, \citenamefont {Govender}, \citenamefont
  {Griggio}, \citenamefont {Grigoryan}, \citenamefont {Gronin}, \citenamefont
  {Gukelberger}, \citenamefont {Heedt}, \citenamefont {Zamorano}, \citenamefont
  {Ho}, \citenamefont {Holgaard}, \citenamefont {Nielsen}, \citenamefont
  {Ingerslev}, \citenamefont {Krogstrup}, \citenamefont {Johansson},
  \citenamefont {Jones}, \citenamefont {Kallaher}, \citenamefont {Karimi},
  \citenamefont {Karzig}, \citenamefont {King}, \citenamefont {Kloster},
  \citenamefont {Knapp}, \citenamefont {Kocon}, \citenamefont {Koski},
  \citenamefont {Kostamo}, \citenamefont {Kumar}, \citenamefont {Laeven},
  \citenamefont {Larsen}, \citenamefont {Li}, \citenamefont {Lindemann},
  \citenamefont {Love}, \citenamefont {Lutchyn}, \citenamefont {Manfra},
  \citenamefont {Memisevic}, \citenamefont {Nayak}, \citenamefont {Nijholt},
  \citenamefont {Madsen}, \citenamefont {Markussen}, \citenamefont {Martinez},
  \citenamefont {McNeil}, \citenamefont {Mullally}, \citenamefont {Nielsen},
  \citenamefont {Nurmohamed}, \citenamefont {O'Farrell}, \citenamefont {Otani},
  \citenamefont {Pauka}, \citenamefont {Petersson}, \citenamefont {Petit},
  \citenamefont {Pikulin}, \citenamefont {Preiss}, \citenamefont {Perez},
  \citenamefont {Rasmussen}, \citenamefont {Rajpalke}, \citenamefont
  {Razmadze}, \citenamefont {Reentila}, \citenamefont {Reilly}, \citenamefont
  {Rouse}, \citenamefont {Sadovskyy}, \citenamefont {Sainiemi}, \citenamefont
  {Schreppler}, \citenamefont {Sidorkin}, \citenamefont {Singh}, \citenamefont
  {Singh}, \citenamefont {Sinha}, \citenamefont {Sohr}, \citenamefont
  {Stankevič}, \citenamefont {Stek}, \citenamefont {Suominen}, \citenamefont
  {Suter}, \citenamefont {Svidenko}, \citenamefont {Teicher}, \citenamefont
  {Temuerhan}, \citenamefont {Thiyagarajah}, \citenamefont {Tholapi},
  \citenamefont {Thomas}, \citenamefont {Toomey}, \citenamefont {Upadhyay},
  \citenamefont {Urban}, \citenamefont {Vaitiekėnas}, \citenamefont
  {Van~Hoogdalem}, \citenamefont {Viazmitinov}, \citenamefont {Waddy},
  \citenamefont {Van~Woerkom}, \citenamefont {Vogel}, \citenamefont {Watson},
  \citenamefont {Weston}, \citenamefont {Winkler}, \citenamefont {Yang},
  \citenamefont {Yau}, \citenamefont {Yi}, \citenamefont {Yucelen},
  \citenamefont {Webster}, \citenamefont {Zeisel},\ and\ \citenamefont
  {Zhao}}]{https://doi.org/10.48550/arxiv.2207.02472}%
  \BibitemOpen
  \bibfield  {author} {\bibinfo {author} {\bibfnamefont {M.}~\bibnamefont
  {Aghaee}}, \bibinfo {author} {\bibfnamefont {A.}~\bibnamefont {Akkala}},
  \bibinfo {author} {\bibfnamefont {Z.}~\bibnamefont {Alam}}, \bibinfo {author}
  {\bibfnamefont {R.}~\bibnamefont {Ali}}, \bibinfo {author} {\bibfnamefont
  {A.~A.}\ \bibnamefont {Ramirez}}, \bibinfo {author} {\bibfnamefont
  {M.}~\bibnamefont {Andrzejczuk}}, \bibinfo {author} {\bibfnamefont {A.~E.}\
  \bibnamefont {Antipov}}, \bibinfo {author} {\bibfnamefont {P.}~\bibnamefont
  {Aseev}}, \bibinfo {author} {\bibfnamefont {M.}~\bibnamefont {Astafev}},
  \bibinfo {author} {\bibfnamefont {B.}~\bibnamefont {Bauer}}, \bibinfo
  {author} {\bibfnamefont {J.}~\bibnamefont {Becker}}, \bibinfo {author}
  {\bibfnamefont {S.}~\bibnamefont {Boddapati}}, \bibinfo {author}
  {\bibfnamefont {F.}~\bibnamefont {Boekhout}}, \bibinfo {author}
  {\bibfnamefont {J.}~\bibnamefont {Bommer}}, \bibinfo {author} {\bibfnamefont
  {E.~B.}\ \bibnamefont {Hansen}}, \bibinfo {author} {\bibfnamefont
  {T.}~\bibnamefont {Bosma}}, \bibinfo {author} {\bibfnamefont
  {L.}~\bibnamefont {Bourdet}}, \bibinfo {author} {\bibfnamefont
  {S.}~\bibnamefont {Boutin}}, \bibinfo {author} {\bibfnamefont
  {P.}~\bibnamefont {Caroff}}, \bibinfo {author} {\bibfnamefont
  {L.}~\bibnamefont {Casparis}}, \bibinfo {author} {\bibfnamefont
  {M.}~\bibnamefont {Cassidy}}, \bibinfo {author} {\bibfnamefont {A.~W.}\
  \bibnamefont {Christensen}}, \bibinfo {author} {\bibfnamefont
  {N.}~\bibnamefont {Clay}}, \bibinfo {author} {\bibfnamefont {W.~S.}\
  \bibnamefont {Cole}}, \bibinfo {author} {\bibfnamefont {F.}~\bibnamefont
  {Corsetti}}, \bibinfo {author} {\bibfnamefont {A.}~\bibnamefont {Cui}},
  \bibinfo {author} {\bibfnamefont {P.}~\bibnamefont {Dalampiras}}, \bibinfo
  {author} {\bibfnamefont {A.}~\bibnamefont {Dokania}}, \bibinfo {author}
  {\bibfnamefont {G.}~\bibnamefont {de~Lange}}, \bibinfo {author}
  {\bibfnamefont {M.}~\bibnamefont {de~Moor}}, \bibinfo {author} {\bibfnamefont
  {J.~C.~E.}\ \bibnamefont {Saldaña}}, \bibinfo {author} {\bibfnamefont
  {S.}~\bibnamefont {Fallahi}}, \bibinfo {author} {\bibfnamefont {Z.~H.}\
  \bibnamefont {Fathabad}}, \bibinfo {author} {\bibfnamefont {J.}~\bibnamefont
  {Gamble}}, \bibinfo {author} {\bibfnamefont {G.}~\bibnamefont {Gardner}},
  \bibinfo {author} {\bibfnamefont {D.}~\bibnamefont {Govender}}, \bibinfo
  {author} {\bibfnamefont {F.}~\bibnamefont {Griggio}}, \bibinfo {author}
  {\bibfnamefont {R.}~\bibnamefont {Grigoryan}}, \bibinfo {author}
  {\bibfnamefont {S.}~\bibnamefont {Gronin}}, \bibinfo {author} {\bibfnamefont
  {J.}~\bibnamefont {Gukelberger}}, \bibinfo {author} {\bibfnamefont
  {S.}~\bibnamefont {Heedt}}, \bibinfo {author} {\bibfnamefont {J.~H.}\
  \bibnamefont {Zamorano}}, \bibinfo {author} {\bibfnamefont {S.}~\bibnamefont
  {Ho}}, \bibinfo {author} {\bibfnamefont {U.~L.}\ \bibnamefont {Holgaard}},
  \bibinfo {author} {\bibfnamefont {W.~H.~P.}\ \bibnamefont {Nielsen}},
  \bibinfo {author} {\bibfnamefont {H.}~\bibnamefont {Ingerslev}}, \bibinfo
  {author} {\bibfnamefont {P.~J.}\ \bibnamefont {Krogstrup}}, \bibinfo {author}
  {\bibfnamefont {L.}~\bibnamefont {Johansson}}, \bibinfo {author}
  {\bibfnamefont {J.}~\bibnamefont {Jones}}, \bibinfo {author} {\bibfnamefont
  {R.}~\bibnamefont {Kallaher}}, \bibinfo {author} {\bibfnamefont
  {F.}~\bibnamefont {Karimi}}, \bibinfo {author} {\bibfnamefont
  {T.}~\bibnamefont {Karzig}}, \bibinfo {author} {\bibfnamefont
  {C.}~\bibnamefont {King}}, \bibinfo {author} {\bibfnamefont {M.~E.}\
  \bibnamefont {Kloster}}, \bibinfo {author} {\bibfnamefont {C.}~\bibnamefont
  {Knapp}}, \bibinfo {author} {\bibfnamefont {D.}~\bibnamefont {Kocon}},
  \bibinfo {author} {\bibfnamefont {J.}~\bibnamefont {Koski}}, \bibinfo
  {author} {\bibfnamefont {P.}~\bibnamefont {Kostamo}}, \bibinfo {author}
  {\bibfnamefont {M.}~\bibnamefont {Kumar}}, \bibinfo {author} {\bibfnamefont
  {T.}~\bibnamefont {Laeven}}, \bibinfo {author} {\bibfnamefont
  {T.}~\bibnamefont {Larsen}}, \bibinfo {author} {\bibfnamefont
  {K.}~\bibnamefont {Li}}, \bibinfo {author} {\bibfnamefont {T.}~\bibnamefont
  {Lindemann}}, \bibinfo {author} {\bibfnamefont {J.}~\bibnamefont {Love}},
  \bibinfo {author} {\bibfnamefont {R.}~\bibnamefont {Lutchyn}}, \bibinfo
  {author} {\bibfnamefont {M.}~\bibnamefont {Manfra}}, \bibinfo {author}
  {\bibfnamefont {E.}~\bibnamefont {Memisevic}}, \bibinfo {author}
  {\bibfnamefont {C.}~\bibnamefont {Nayak}}, \bibinfo {author} {\bibfnamefont
  {B.}~\bibnamefont {Nijholt}}, \bibinfo {author} {\bibfnamefont {M.~H.}\
  \bibnamefont {Madsen}}, \bibinfo {author} {\bibfnamefont {S.}~\bibnamefont
  {Markussen}}, \bibinfo {author} {\bibfnamefont {E.}~\bibnamefont {Martinez}},
  \bibinfo {author} {\bibfnamefont {R.}~\bibnamefont {McNeil}}, \bibinfo
  {author} {\bibfnamefont {A.}~\bibnamefont {Mullally}}, \bibinfo {author}
  {\bibfnamefont {J.}~\bibnamefont {Nielsen}}, \bibinfo {author} {\bibfnamefont
  {A.}~\bibnamefont {Nurmohamed}}, \bibinfo {author} {\bibfnamefont
  {E.}~\bibnamefont {O'Farrell}}, \bibinfo {author} {\bibfnamefont
  {K.}~\bibnamefont {Otani}}, \bibinfo {author} {\bibfnamefont
  {S.}~\bibnamefont {Pauka}}, \bibinfo {author} {\bibfnamefont
  {K.}~\bibnamefont {Petersson}}, \bibinfo {author} {\bibfnamefont
  {L.}~\bibnamefont {Petit}}, \bibinfo {author} {\bibfnamefont
  {D.}~\bibnamefont {Pikulin}}, \bibinfo {author} {\bibfnamefont
  {F.}~\bibnamefont {Preiss}}, \bibinfo {author} {\bibfnamefont {M.~Q.}\
  \bibnamefont {Perez}}, \bibinfo {author} {\bibfnamefont {K.}~\bibnamefont
  {Rasmussen}}, \bibinfo {author} {\bibfnamefont {M.}~\bibnamefont {Rajpalke}},
  \bibinfo {author} {\bibfnamefont {D.}~\bibnamefont {Razmadze}}, \bibinfo
  {author} {\bibfnamefont {O.}~\bibnamefont {Reentila}}, \bibinfo {author}
  {\bibfnamefont {D.}~\bibnamefont {Reilly}}, \bibinfo {author} {\bibfnamefont
  {R.}~\bibnamefont {Rouse}}, \bibinfo {author} {\bibfnamefont
  {I.}~\bibnamefont {Sadovskyy}}, \bibinfo {author} {\bibfnamefont
  {L.}~\bibnamefont {Sainiemi}}, \bibinfo {author} {\bibfnamefont
  {S.}~\bibnamefont {Schreppler}}, \bibinfo {author} {\bibfnamefont
  {V.}~\bibnamefont {Sidorkin}}, \bibinfo {author} {\bibfnamefont
  {A.}~\bibnamefont {Singh}}, \bibinfo {author} {\bibfnamefont
  {S.}~\bibnamefont {Singh}}, \bibinfo {author} {\bibfnamefont
  {S.}~\bibnamefont {Sinha}}, \bibinfo {author} {\bibfnamefont
  {P.}~\bibnamefont {Sohr}}, \bibinfo {author} {\bibfnamefont {T.}~\bibnamefont
  {Stankevič}}, \bibinfo {author} {\bibfnamefont {L.}~\bibnamefont {Stek}},
  \bibinfo {author} {\bibfnamefont {H.}~\bibnamefont {Suominen}}, \bibinfo
  {author} {\bibfnamefont {J.}~\bibnamefont {Suter}}, \bibinfo {author}
  {\bibfnamefont {V.}~\bibnamefont {Svidenko}}, \bibinfo {author}
  {\bibfnamefont {S.}~\bibnamefont {Teicher}}, \bibinfo {author} {\bibfnamefont
  {M.}~\bibnamefont {Temuerhan}}, \bibinfo {author} {\bibfnamefont
  {N.}~\bibnamefont {Thiyagarajah}}, \bibinfo {author} {\bibfnamefont
  {R.}~\bibnamefont {Tholapi}}, \bibinfo {author} {\bibfnamefont
  {M.}~\bibnamefont {Thomas}}, \bibinfo {author} {\bibfnamefont
  {E.}~\bibnamefont {Toomey}}, \bibinfo {author} {\bibfnamefont
  {S.}~\bibnamefont {Upadhyay}}, \bibinfo {author} {\bibfnamefont
  {I.}~\bibnamefont {Urban}}, \bibinfo {author} {\bibfnamefont
  {S.}~\bibnamefont {Vaitiekėnas}}, \bibinfo {author} {\bibfnamefont
  {K.}~\bibnamefont {Van~Hoogdalem}}, \bibinfo {author} {\bibfnamefont {D.~V.}\
  \bibnamefont {Viazmitinov}}, \bibinfo {author} {\bibfnamefont
  {S.}~\bibnamefont {Waddy}}, \bibinfo {author} {\bibfnamefont
  {D.}~\bibnamefont {Van~Woerkom}}, \bibinfo {author} {\bibfnamefont
  {D.}~\bibnamefont {Vogel}}, \bibinfo {author} {\bibfnamefont
  {J.}~\bibnamefont {Watson}}, \bibinfo {author} {\bibfnamefont
  {J.}~\bibnamefont {Weston}}, \bibinfo {author} {\bibfnamefont {G.~W.}\
  \bibnamefont {Winkler}}, \bibinfo {author} {\bibfnamefont {C.~K.}\
  \bibnamefont {Yang}}, \bibinfo {author} {\bibfnamefont {S.}~\bibnamefont
  {Yau}}, \bibinfo {author} {\bibfnamefont {D.}~\bibnamefont {Yi}}, \bibinfo
  {author} {\bibfnamefont {E.}~\bibnamefont {Yucelen}}, \bibinfo {author}
  {\bibfnamefont {A.}~\bibnamefont {Webster}}, \bibinfo {author} {\bibfnamefont
  {R.}~\bibnamefont {Zeisel}},\ and\ \bibinfo {author} {\bibfnamefont
  {R.}~\bibnamefont {Zhao}},\ }\href
  {https://doi.org/10.48550/ARXIV.2207.02472} {\bibinfo {title} {Inas-al hybrid
  devices passing the topological gap protocol}} (\bibinfo {year}
  {2022})\BibitemShut {NoStop}%
\bibitem [{\citenamefont {Kejriwal}\ and\ \citenamefont
  {Muralidharan}(2022)}]{kejriwal2022}%
  \BibitemOpen
  \bibfield  {author} {\bibinfo {author} {\bibfnamefont {A.}~\bibnamefont
  {Kejriwal}}\ and\ \bibinfo {author} {\bibfnamefont {B.}~\bibnamefont
  {Muralidharan}},\ }\href {https://doi.org/10.1103/PhysRevB.105.L161403}
  {\bibfield  {journal} {\bibinfo  {journal} {Phys. Rev. B}\ }\textbf {\bibinfo
  {volume} {105}},\ \bibinfo {pages} {L161403} (\bibinfo {year}
  {2022})}\BibitemShut {NoStop}%
\bibitem [{\citenamefont {Pan}\ \emph {et~al.}(2021{\natexlab{c}})\citenamefont
  {Pan}, \citenamefont {Sau},\ and\ \citenamefont {Das~Sarma}}]{Pan_nonlocal}%
  \BibitemOpen
  \bibfield  {author} {\bibinfo {author} {\bibfnamefont {H.}~\bibnamefont
  {Pan}}, \bibinfo {author} {\bibfnamefont {J.~D.}\ \bibnamefont {Sau}},\ and\
  \bibinfo {author} {\bibfnamefont {S.}~\bibnamefont {Das~Sarma}},\ }\href
  {https://doi.org/10.1103/PhysRevB.103.014513} {\bibfield  {journal} {\bibinfo
   {journal} {Phys. Rev. B}\ }\textbf {\bibinfo {volume} {103}},\ \bibinfo
  {pages} {014513} (\bibinfo {year} {2021}{\natexlab{c}})}\BibitemShut
  {NoStop}%
\bibitem [{\citenamefont {Ryu}\ and\ \citenamefont
  {Hatsugai}(2006)}]{PhysRevB.73.245115}%
  \BibitemOpen
  \bibfield  {author} {\bibinfo {author} {\bibfnamefont {S.}~\bibnamefont
  {Ryu}}\ and\ \bibinfo {author} {\bibfnamefont {Y.}~\bibnamefont {Hatsugai}},\
  }\href {https://doi.org/10.1103/PhysRevB.73.245115} {\bibfield  {journal}
  {\bibinfo  {journal} {Phys. Rev. B}\ }\textbf {\bibinfo {volume} {73}},\
  \bibinfo {pages} {245115} (\bibinfo {year} {2006})}\BibitemShut {NoStop}%
\bibitem [{\citenamefont {Vidal}\ \emph {et~al.}(2003)\citenamefont {Vidal},
  \citenamefont {Latorre}, \citenamefont {Rico},\ and\ \citenamefont
  {Kitaev}}]{PhysRevLett.90.227902}%
  \BibitemOpen
  \bibfield  {author} {\bibinfo {author} {\bibfnamefont {G.}~\bibnamefont
  {Vidal}}, \bibinfo {author} {\bibfnamefont {J.~I.}\ \bibnamefont {Latorre}},
  \bibinfo {author} {\bibfnamefont {E.}~\bibnamefont {Rico}},\ and\ \bibinfo
  {author} {\bibfnamefont {A.}~\bibnamefont {Kitaev}},\ }\href
  {https://doi.org/10.1103/PhysRevLett.90.227902} {\bibfield  {journal}
  {\bibinfo  {journal} {Phys. Rev. Lett.}\ }\textbf {\bibinfo {volume} {90}},\
  \bibinfo {pages} {227902} (\bibinfo {year} {2003})}\BibitemShut {NoStop}%
\bibitem [{\citenamefont {Koffel}\ \emph {et~al.}(2012)\citenamefont {Koffel},
  \citenamefont {Lewenstein},\ and\ \citenamefont
  {Tagliacozzo}}]{PhysRevLett.109.267203}%
  \BibitemOpen
  \bibfield  {author} {\bibinfo {author} {\bibfnamefont {T.}~\bibnamefont
  {Koffel}}, \bibinfo {author} {\bibfnamefont {M.}~\bibnamefont {Lewenstein}},\
  and\ \bibinfo {author} {\bibfnamefont {L.}~\bibnamefont {Tagliacozzo}},\
  }\href {https://doi.org/10.1103/PhysRevLett.109.267203} {\bibfield  {journal}
  {\bibinfo  {journal} {Phys. Rev. Lett.}\ }\textbf {\bibinfo {volume} {109}},\
  \bibinfo {pages} {267203} (\bibinfo {year} {2012})}\BibitemShut {NoStop}%
\bibitem [{\citenamefont {Fromholz}\ \emph {et~al.}(2020)\citenamefont
  {Fromholz}, \citenamefont {Magnifico}, \citenamefont {Vitale}, \citenamefont
  {Mendes-Santos},\ and\ \citenamefont {Dalmonte}}]{PhysRevB.101.085136}%
  \BibitemOpen
  \bibfield  {author} {\bibinfo {author} {\bibfnamefont {P.}~\bibnamefont
  {Fromholz}}, \bibinfo {author} {\bibfnamefont {G.}~\bibnamefont {Magnifico}},
  \bibinfo {author} {\bibfnamefont {V.}~\bibnamefont {Vitale}}, \bibinfo
  {author} {\bibfnamefont {T.}~\bibnamefont {Mendes-Santos}},\ and\ \bibinfo
  {author} {\bibfnamefont {M.}~\bibnamefont {Dalmonte}},\ }\href
  {https://doi.org/10.1103/PhysRevB.101.085136} {\bibfield  {journal} {\bibinfo
   {journal} {Phys. Rev. B}\ }\textbf {\bibinfo {volume} {101}},\ \bibinfo
  {pages} {085136} (\bibinfo {year} {2020})}\BibitemShut {NoStop}%
\bibitem [{\citenamefont {Zeng}\ \emph {et~al.}(2019)\citenamefont {Zeng},
  \citenamefont {Chen}, \citenamefont {Zhou},\ and\ \citenamefont
  {Wen}}]{Zeng2019}%
  \BibitemOpen
  \bibfield  {author} {\bibinfo {author} {\bibfnamefont {B.}~\bibnamefont
  {Zeng}}, \bibinfo {author} {\bibfnamefont {X.}~\bibnamefont {Chen}}, \bibinfo
  {author} {\bibfnamefont {D.-L.}\ \bibnamefont {Zhou}},\ and\ \bibinfo
  {author} {\bibfnamefont {X.-G.}\ \bibnamefont {Wen}},\ }\bibinfo {title}
  {Correlation and entanglement},\ in\ \href
  {https://doi.org/10.1007/978-1-4939-9084-9_1} {\emph {\bibinfo {booktitle}
  {Quantum Information Meets Quantum Matter: From Quantum Entanglement to
  Topological Phases of Many-Body Systems}}}\ (\bibinfo  {publisher} {Springer
  New York},\ \bibinfo {address} {New York, NY},\ \bibinfo {year} {2019})\ pp.\
  \bibinfo {pages} {3--35}\BibitemShut {NoStop}%
\bibitem [{\citenamefont {Micallo}\ \emph {et~al.}(2020)\citenamefont
  {Micallo}, \citenamefont {Vitale}, \citenamefont {Dalmonte},\ and\
  \citenamefont {Fromholz}}]{10.21468/SciPostPhysCore.3.2.012}%
  \BibitemOpen
  \bibfield  {author} {\bibinfo {author} {\bibfnamefont {T.}~\bibnamefont
  {Micallo}}, \bibinfo {author} {\bibfnamefont {V.}~\bibnamefont {Vitale}},
  \bibinfo {author} {\bibfnamefont {M.}~\bibnamefont {Dalmonte}},\ and\
  \bibinfo {author} {\bibfnamefont {P.}~\bibnamefont {Fromholz}},\ }\href
  {https://doi.org/10.21468/SciPostPhysCore.3.2.012} {\bibfield  {journal}
  {\bibinfo  {journal} {SciPost Phys. Core}\ }\textbf {\bibinfo {volume} {3}},\
  \bibinfo {pages} {12} (\bibinfo {year} {2020})}\BibitemShut {NoStop}%
\bibitem [{\citenamefont {Bianchi}\ \emph {et~al.}(2022)\citenamefont
  {Bianchi}, \citenamefont {Hackl}, \citenamefont {Kieburg}, \citenamefont
  {Rigol},\ and\ \citenamefont {Vidmar}}]{Bianchi-volume}%
  \BibitemOpen
  \bibfield  {author} {\bibinfo {author} {\bibfnamefont {E.}~\bibnamefont
  {Bianchi}}, \bibinfo {author} {\bibfnamefont {L.}~\bibnamefont {Hackl}},
  \bibinfo {author} {\bibfnamefont {M.}~\bibnamefont {Kieburg}}, \bibinfo
  {author} {\bibfnamefont {M.}~\bibnamefont {Rigol}},\ and\ \bibinfo {author}
  {\bibfnamefont {L.}~\bibnamefont {Vidmar}},\ }\href
  {https://doi.org/10.1103/PRXQuantum.3.030201} {\bibfield  {journal} {\bibinfo
   {journal} {PRX Quantum}\ }\textbf {\bibinfo {volume} {3}},\ \bibinfo {pages}
  {030201} (\bibinfo {year} {2022})}\BibitemShut {NoStop}%
\bibitem [{\citenamefont {P\"oyh\"onen}\ \emph {et~al.}(2022)\citenamefont
  {P\"oyh\"onen}, \citenamefont {Moghaddam},\ and\ \citenamefont
  {Ojanen}}]{uncertainity_finland}%
  \BibitemOpen
  \bibfield  {author} {\bibinfo {author} {\bibfnamefont {K.}~\bibnamefont
  {P\"oyh\"onen}}, \bibinfo {author} {\bibfnamefont {A.~G.}\ \bibnamefont
  {Moghaddam}},\ and\ \bibinfo {author} {\bibfnamefont {T.}~\bibnamefont
  {Ojanen}},\ }\href {https://doi.org/10.1103/PhysRevResearch.4.023200}
  {\bibfield  {journal} {\bibinfo  {journal} {Phys. Rev. Res.}\ }\textbf
  {\bibinfo {volume} {4}},\ \bibinfo {pages} {023200} (\bibinfo {year}
  {2022})}\BibitemShut {NoStop}%
\bibitem [{\citenamefont {Pan}\ \emph {et~al.}(2019)\citenamefont {Pan},
  \citenamefont {Sau}, \citenamefont {Stanescu},\ and\ \citenamefont
  {Das~Sarma}}]{Pan-2019-prb}%
  \BibitemOpen
  \bibfield  {author} {\bibinfo {author} {\bibfnamefont {H.}~\bibnamefont
  {Pan}}, \bibinfo {author} {\bibfnamefont {J.~D.}\ \bibnamefont {Sau}},
  \bibinfo {author} {\bibfnamefont {T.~D.}\ \bibnamefont {Stanescu}},\ and\
  \bibinfo {author} {\bibfnamefont {S.}~\bibnamefont {Das~Sarma}},\ }\href
  {https://doi.org/10.1103/PhysRevB.99.054507} {\bibfield  {journal} {\bibinfo
  {journal} {Phys. Rev. B}\ }\textbf {\bibinfo {volume} {99}},\ \bibinfo
  {pages} {054507} (\bibinfo {year} {2019})}\BibitemShut {NoStop}%
\bibitem [{\citenamefont {Pan}\ and\ \citenamefont {Sarma}(2020)}]{Pan-2020}%
  \BibitemOpen
  \bibfield  {author} {\bibinfo {author} {\bibfnamefont {H.}~\bibnamefont
  {Pan}}\ and\ \bibinfo {author} {\bibfnamefont {S.~D.}\ \bibnamefont
  {Sarma}},\ }\href@noop {} {\bibfield  {journal} {\bibinfo  {journal} {Phys.
  Rev. Research}\ }\textbf {\bibinfo {volume} {2}},\ \bibinfo {pages} {013377}
  (\bibinfo {year} {2020})}\BibitemShut {NoStop}%
\bibitem [{\citenamefont {Moore}\ \emph {et~al.}(2018)\citenamefont {Moore},
  \citenamefont {Stanescu},\ and\ \citenamefont {Tewari}}]{PhysRevB.97.165302}%
  \BibitemOpen
  \bibfield  {author} {\bibinfo {author} {\bibfnamefont {C.}~\bibnamefont
  {Moore}}, \bibinfo {author} {\bibfnamefont {T.~D.}\ \bibnamefont
  {Stanescu}},\ and\ \bibinfo {author} {\bibfnamefont {S.}~\bibnamefont
  {Tewari}},\ }\href {https://doi.org/10.1103/PhysRevB.97.165302} {\bibfield
  {journal} {\bibinfo  {journal} {Phys. Rev. B}\ }\textbf {\bibinfo {volume}
  {97}},\ \bibinfo {pages} {165302} (\bibinfo {year} {2018})}\BibitemShut
  {NoStop}%
\bibitem [{\citenamefont {Chen}\ \emph {et~al.}(2019)\citenamefont {Chen},
  \citenamefont {Woods}, \citenamefont {Yu}, \citenamefont {Hocevar},
  \citenamefont {Car}, \citenamefont {Plissard}, \citenamefont {Bakkers},
  \citenamefont {Stanescu},\ and\ \citenamefont
  {Frolov}}]{PhysRevLett.123.107703}%
  \BibitemOpen
  \bibfield  {author} {\bibinfo {author} {\bibfnamefont {J.}~\bibnamefont
  {Chen}}, \bibinfo {author} {\bibfnamefont {B.~D.}\ \bibnamefont {Woods}},
  \bibinfo {author} {\bibfnamefont {P.}~\bibnamefont {Yu}}, \bibinfo {author}
  {\bibfnamefont {M.}~\bibnamefont {Hocevar}}, \bibinfo {author} {\bibfnamefont
  {D.}~\bibnamefont {Car}}, \bibinfo {author} {\bibfnamefont {S.~R.}\
  \bibnamefont {Plissard}}, \bibinfo {author} {\bibfnamefont {E.~P. A.~M.}\
  \bibnamefont {Bakkers}}, \bibinfo {author} {\bibfnamefont {T.~D.}\
  \bibnamefont {Stanescu}},\ and\ \bibinfo {author} {\bibfnamefont {S.~M.}\
  \bibnamefont {Frolov}},\ }\href
  {https://doi.org/10.1103/PhysRevLett.123.107703} {\bibfield  {journal}
  {\bibinfo  {journal} {Phys. Rev. Lett.}\ }\textbf {\bibinfo {volume} {123}},\
  \bibinfo {pages} {107703} (\bibinfo {year} {2019})}\BibitemShut {NoStop}%
\bibitem [{\citenamefont {Deng}\ \emph
  {et~al.}(2016{\natexlab{b}})\citenamefont {Deng}, \citenamefont
  {Vaitiekėnas}, \citenamefont {Hansen}, \citenamefont {Danon}, \citenamefont
  {Leijnse}, \citenamefont {Flensberg}, \citenamefont {Nygård}, \citenamefont
  {Krogstrup},\ and\ \citenamefont {Marcus}}]{doi:10.1126/science.aaf3961}%
  \BibitemOpen
  \bibfield  {author} {\bibinfo {author} {\bibfnamefont {M.~T.}\ \bibnamefont
  {Deng}}, \bibinfo {author} {\bibfnamefont {S.}~\bibnamefont {Vaitiekėnas}},
  \bibinfo {author} {\bibfnamefont {E.~B.}\ \bibnamefont {Hansen}}, \bibinfo
  {author} {\bibfnamefont {J.}~\bibnamefont {Danon}}, \bibinfo {author}
  {\bibfnamefont {M.}~\bibnamefont {Leijnse}}, \bibinfo {author} {\bibfnamefont
  {K.}~\bibnamefont {Flensberg}}, \bibinfo {author} {\bibfnamefont
  {J.}~\bibnamefont {Nygård}}, \bibinfo {author} {\bibfnamefont
  {P.}~\bibnamefont {Krogstrup}},\ and\ \bibinfo {author} {\bibfnamefont
  {C.~M.}\ \bibnamefont {Marcus}},\ }\href
  {https://doi.org/10.1126/science.aaf3961} {\bibfield  {journal} {\bibinfo
  {journal} {Science}\ }\textbf {\bibinfo {volume} {354}},\ \bibinfo {pages}
  {1557} (\bibinfo {year} {2016}{\natexlab{b}})},\ \Eprint
  {https://arxiv.org/abs/https://www.science.org/doi/pdf/10.1126/science.aaf3961}
  {https://www.science.org/doi/pdf/10.1126/science.aaf3961} \BibitemShut
  {NoStop}%
\bibitem [{\citenamefont {Duse}\ \emph {et~al.}(2021)\citenamefont {Duse},
  \citenamefont {Sriram}, \citenamefont {Gharavi}, \citenamefont {Baugh},\ and\
  \citenamefont {Muralidharan}}]{Duse_2021}%
  \BibitemOpen
  \bibfield  {author} {\bibinfo {author} {\bibfnamefont {C.}~\bibnamefont
  {Duse}}, \bibinfo {author} {\bibfnamefont {P.}~\bibnamefont {Sriram}},
  \bibinfo {author} {\bibfnamefont {K.}~\bibnamefont {Gharavi}}, \bibinfo
  {author} {\bibfnamefont {J.}~\bibnamefont {Baugh}},\ and\ \bibinfo {author}
  {\bibfnamefont {B.}~\bibnamefont {Muralidharan}},\ }\href
  {https://doi.org/10.1088/1361-648x/ac0d16} {\bibfield  {journal} {\bibinfo
  {journal} {Journal of Physics: Condensed Matter}\ }\textbf {\bibinfo {volume}
  {33}},\ \bibinfo {pages} {365301} (\bibinfo {year} {2021})}\BibitemShut
  {NoStop}%
\bibitem [{\citenamefont {Cayao}(2017)}]{thesis_superconducting_floquet}%
  \BibitemOpen
  \bibfield  {author} {\bibinfo {author} {\bibfnamefont {J.}~\bibnamefont
  {Cayao}},\ }\href {https://doi.org/10.48550/ARXIV.1703.07630} {\bibinfo
  {title} {Hybrid superconductor-semiconductor nanowire junctions as useful
  platforms to study majorana bound states}} (\bibinfo {year}
  {2017})\BibitemShut {NoStop}%
\bibitem [{sup()}]{supple}%
  \BibitemOpen
  \href@noop {} {\bibinfo {title} {See supplemental material for the expanded
  discussion of the entanglement entropy calculation as well as the quantum
  transport calculations of the local and nonlocal conductances}}\BibitemShut
  {NoStop}%
\bibitem [{\citenamefont {Kitaev}\ and\ \citenamefont
  {Preskill}(2006)}]{TEE-kitaev}%
  \BibitemOpen
  \bibfield  {author} {\bibinfo {author} {\bibfnamefont {A.}~\bibnamefont
  {Kitaev}}\ and\ \bibinfo {author} {\bibfnamefont {J.}~\bibnamefont
  {Preskill}},\ }\href {https://doi.org/10.1103/PhysRevLett.96.110404}
  {\bibfield  {journal} {\bibinfo  {journal} {Phys. Rev. Lett.}\ }\textbf
  {\bibinfo {volume} {96}},\ \bibinfo {pages} {110404} (\bibinfo {year}
  {2006})}\BibitemShut {NoStop}%
\bibitem [{\citenamefont {San-Jose}\ \emph {et~al.}(2013)\citenamefont
  {San-Jose}, \citenamefont {Cayao}, \citenamefont {Prada},\ and\ \citenamefont
  {Aguado}}]{San-Jose_2013}%
  \BibitemOpen
  \bibfield  {author} {\bibinfo {author} {\bibfnamefont {P.}~\bibnamefont
  {San-Jose}}, \bibinfo {author} {\bibfnamefont {J.}~\bibnamefont {Cayao}},
  \bibinfo {author} {\bibfnamefont {E.}~\bibnamefont {Prada}},\ and\ \bibinfo
  {author} {\bibfnamefont {R.}~\bibnamefont {Aguado}},\ }\href
  {https://doi.org/10.1088/1367-2630/15/7/075019} {\bibfield  {journal}
  {\bibinfo  {journal} {New Journal of Physics}\ }\textbf {\bibinfo {volume}
  {15}},\ \bibinfo {pages} {075019} (\bibinfo {year} {2013})}\BibitemShut
  {NoStop}%
\end{thebibliography}%

% %\end{acknowledgements}
% \bibliography{main}

\end{document}